\documentclass[12pt]{iopart}

\usepackage{epsfig,afterpage}
\usepackage{graphicx}
\usepackage{amsfonts,bbold}
\usepackage{amssymb}
\usepackage{indentfirst}
\usepackage{setstack}
\usepackage{amsthm}
\usepackage{dsfont}
\usepackage{epsfig}
\usepackage{subfigure}
\usepackage{iopams}
\usepackage{multirow}
\usepackage{pstricks}
\usepackage{psfrag}
\usepackage{auto-pst-pdf}

\def\beq{\begin{equation}}
\def\eeq{\end{equation}}
\def\bea{\begin{eqnarray}}
\def\eea{\end{eqnarray}}
\def\bq{\begin{quote}}
\def\eq{\end{quote}}
\def\ben{\begin{enumerate}}
\def\een{\end{enumerate}}
\def\bit{\begin{itemize}}
\def\eit{\end{itemize}}
\def\nn{\nonumber}

\def\eqref{\eref}
\def\acknowledgments{\ack}

\def\ident{\mathbb{1}}

\begin{document}

\title[Dynamics of spin chains with tensor networks]{
Tensor network techniques for the computation of dynamical observables in 1D quantum spin systems
}


%

 \author{Alexander M\"uller-Hermes, J. Ignacio Cirac and Mari Carmen Ba\~nuls}
 \address{Max-Planck-Institut f\"ur Quantenoptik,
   Hans-Kopfermann-Str. 1, 85748 Garching, Germany.}
 \ead{banulsm@mpq.mpg.de} 
 
\begin{abstract}
We analyze the recently developed folding algorithm [Phys. Rev. Lett. 102, 240603 (2009)] 
to simulate the dynamics of infinite quantum spin chains, and relate its performance to the kind of 
entanglement  produced under the evolution of product states. We benchmark the accomplishments of this technique with respect to alternative strategies using Ising Hamiltonians with transverse and parallel fields, as well as XY models. Additionally, we evaluate its ability to find ground and thermal equilibrium states.
\end{abstract}

\pacs{03.67.Mn, 02.70.-c, 75.10.Pq}

\maketitle

\section{Introduction}

Numerical techniques are fundamental for the description of quantum many body physics,
since exactly solvable systems are the exception and analytic
approximations have a limited range of validity.
But given the exponential growth of the Hilbert space dimension with the number of constituents 
the exact numerical solution is affordable for only small system sizes, 
so that approximate algorithms are in general required.
While different techniques exist, including quantum Monte Carlo methods, density functional theory, and 
various tensor network algorithms, to deal with the equilibrium properties of such systems, their 
applicability to out of equilibrium situations is typically limited.

Currently the most prominent technique for simulating the dynamics of one dimensional systems is the TEBD 
algorithm~\cite{vidal03eff,daley04adapt,vidal07infinite}, based on matrix product states (MPS)~\cite{aklt88,kluemper91,kluemper92,fannes92fcs,verstraete04dmrg,perez07mps}.
But the potentially fast increase of entanglement under out of equilibrium evolution~\cite{calabrese05,schuch08entropy,osborne06efficient}
can make it fail after short times.
It is then highly desirable to develop new alternative methods for time evolution, as well as to understand the range of validity of the different approaches.

Any time evolution, in particular the operation of TEBD, can be understood in terms of tensor networks (TN),
and the computation of time dependent expectation values can be reduced to the problem of a network contraction.
In particular, in the one-dimensional case this network is two-dimensional and TEBD represents 
one possible way of contracting it.
Other contraction strategies exist~\cite{verstraete04mpdo,clark10heis} which may attain some improvement in particular cases.
In all these approaches, the two dimensional TN is approximated (truncated) as one dimensional.
This implies that when the evolution introduces a violation of the area law, i.e. large entanglement along the bonds,
these procedures will incur in a big truncation error.
That is for instance the case in most global quenches, for which these techniques thus face a 
fundamental obstacle.

One way around this, at least for some problems, is performing the TN contraction in a 
different direction, 
with or without first folding the network. 
This may overcome the problem of entanglement, 
as it is not clear that this will fail for all states violating the area law. In~\cite{banuls09fold} we gave evidence 
that, at least for some problems where the entanglement grows maximally, this method performs better than the
techniques mentioned above. 
Nevertheless a more in-depth analysis is required 
to better understand the conditions under which the transverse and folding techniques will be advantageous and when they will also fail.

In this paper we undertake this task. 
First, we construct an explicit tensor network model (a cellular automaton) 
that provides an intuitive understanding of the
properties of the transverse and folding methods, related to the entanglement in
the tensor networks involved. 
In particular, when the initial state contains localized entangled pairs that propagate under the evolution
the area law will be violated due to the increasing entanglement. The folding method is nevertheless able 
to exactly account for this entanglement, whereas more standard techniques are not. 
Second, by studying the analogue to entanglement in the time direction, 
we analyze how closely this idealized picture is reproduced by more realistic spin models.
Third, we describe various dynamical observables that can be computed with the transverse techniques, and 
benchmark their performances for evolution after a global quench and for computing dynamical correlations in the ground state
in the cases of Ising and XY models.
Finally, we also apply the techniques to imaginary time scenarios, and test their outcome for ground and thermal states.

The structure of the paper is the following. In \sref{sec:method} we describe the method in detail, and 
relate it to other alternative TN techniques for the computation of
dynamical observables. \Sref{sec:entanglement} introduces the tensor network model
and analyzes the different kinds of entanglement appearing in the TN, also in the case of an Ising type spin chain.
In \sref{sec:applications} we
enumerate different physical applications for the TN contraction
techniques, and illustrate them with our numerical results for the Ising
and XY chains. Finally, in \sref{sec:conclu} we discuss the
conclusions of this study.

\section{\label{sec:method}The transverse folding method}


\subsection{\label{sec:TNconstruction}Representing time dependent 
observables as a tensor network}

One of the main ideas under this strategy is to restate the problem of
computing a time-dependent observable as that of contracting a tensor network,
without previously approximating the evolved state by a particular TNS.
In particular, we are interested in the evolution under a local Hamiltonian of a 1D system initially described by a MPS.
In the following we describe in detail the construction of the
tensor network for this case.~\footnote{A similar construction can be proposed
  for other initial TNS, although the complexity of the resulting TN and its
  contraction will strongly depend on each particular ansatz.}

The MPS ansatz for a chain of $N$ $d$-dimensional sites has the form
\beq 
|\Psi\rangle =\sum_{i_1,\ldots
  i_N=1}^d \mathrm{tr}(A_1^{i_1}\ldots A_N^{i_N}) |i_1,\ldots i_N
\rangle,
\label{eq:mps}
\eeq where $\{|i\rangle\}_{i=0}^{d-1}$ is the local basis of every site, and
each $A_k^i$ is a $D$-dimensional matrix. The bond dimension,
 $D$, determines the number of parameters in the ansatz.

\begin{figure}[floatfix]
\subfigure[Discretization of time.]{
 \label{fig:fig1-a}
\psfrag{U}[c][l]{$U(t)$}
\psfrag{d}[c][l]{$U(\delta t)$}
  \includegraphics[width=.75\columnwidth]{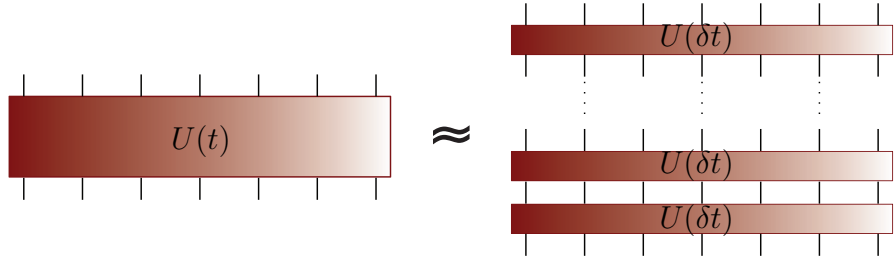}
}
\subfigure[A possible Trotter approximation, to lowest order, for a nearest-neighbour Hamiltonian.]{
 \label{fig:fig1-b}
\psfrag{d}[c][l]{$U(\delta t)$}
\psfrag{e}[c][l]{$e^{-i \delta t H_e}$}
\psfrag{o}[c][l]{$e^{-i \delta t H_o}$}
  \includegraphics[width=.75\columnwidth]{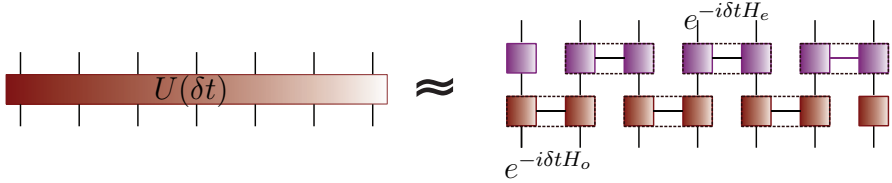}
}
\caption{Scheme of Trotter decomposition of the evolution
  operator. Each two-body term can be written as sum of tensor
  products of single body operators via a singular value
  decomposition. By contracting together the terms for each site, we
  obtain a MPO approximation to the step evolution operator.}
\label{fig:fig1}
\end{figure}

The local Hamiltonian has the form $H=\sum_i h_i$, where
each term $h_i$ acts on only a few neighbouring sites, and the sum runs
over all sites in the chain.  The evolution operator, $U(t)$, that
maps the initial to the final state is in general highly non-local
and thus cannot be directly applied in an efficient
way.  A standard technique to approximate the action of the operator
in a local fashion includes the division of the total time, $t$, in small
discrete steps of length $\delta$ [\fref{fig:fig1-a}]. The evolution operator for 
each step can then be approximated by a local structure, namely a
matrix product operator (MPO)~\cite{murg08mpo} [see
\fref{fig:fig1-b}].  This is usually achieved by a Suzuki-Trotter
decomposition~\cite{trotter59}, in which 
the Hamiltonian is written as a sum of various terms and the exponential 
of the sum is approximated as 
a product of several exponentials of the individual addends.
 The error
of the approximation is
controlled by the time step, $\delta$, and grows linearly with the total time.  The
sum decomposition of the Hamiltonian is not unique
but can be chosen in the most convenient way depending on the problem.
   In particular,
in the case of a nearest-neighbour interaction, $H=\sum_i h_{i,i+1}$, a
general approach is to split the Hamiltonian in a sum of two terms,
each of them containing only mutually commuting operators
$$
H=H_e+H_o\equiv \sum_k h_{2k,2k+1}+\sum_k h_{2k-1,2k}.
$$ Then, each evolution step is given, to the first order in
$\delta$, by $U(\delta) \approx e^{-i \delta H_e} e^{-i \delta H_o}$.
Each of the exponentials above can be exactly written as a MPO which
simply consists of the tensor product of non-overlapping two-body
terms $e^{-i \delta h_{i,i+1}}$ [\fref{fig:fig1-b}]. 
The total evolution operator is then given by a two dimensional tensor
network obtained from the concatenation of all the involved
MPO.~\footnote{In all figures, we use a pictorial representation to
  describe the relevant tensor networks and the related algorithms. In
  this picture, a tensor is represented by a box, with each index
  represented by one leg. A line connecting two tensors represents the
  contraction of a common index between both tensors.}

For some nearest-neighbour Hamiltonians more efficient decompositions exist. It is the case of the two-body term in the Ising Hamiltonian, whose exponential accepts an exact, translationally invariant MPO expression with bond dimension two~\cite{murg08mpo}.
Thus, by decomposing the Ising Hamiltonian in a sum of two-body and single-body terms, it is possible to construct 
an approximate tensor network for the evolution which maintains the translational invariance and results in a more efficient TN decomposition. This is the approach we have used for the numerical simulations presented in this paper.
 
\begin{figure}[floatfix]
\begin{minipage}[t]{.35\columnwidth}
\subfigure[Tensor network description of the evolved state $|\Psi(t)\rangle $.]{
 \label{fig:fig2-a}
\psfrag{U}[c][c]{$U(t)$}
\psfrag{P}[c][c]{$|\Psi\rangle$}
  \includegraphics[width=.95\columnwidth]{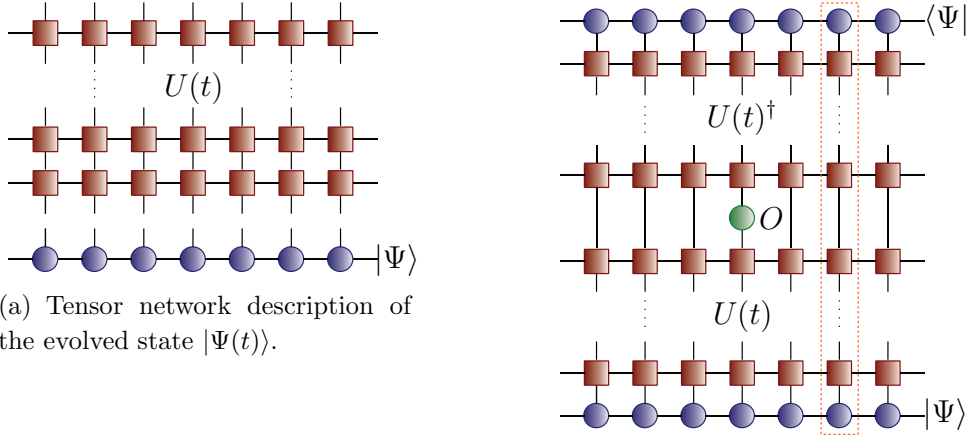}
}
\end{minipage}
\hspace{.1\columnwidth}
\begin{minipage}[t]{.35\columnwidth}
\subfigure[Tensor network representing the time dependent expectation value of a single body operator,$\langle \Psi|O(t)|\Psi\rangle$.]{
 \label{fig:fig2-b}
\psfrag{U}[c][c]{$U(t)$}
\psfrag{V}[c][c]{$U(t)^{\dagger}$}
\psfrag{P}[c][c]{$|\Psi\rangle$}
\psfrag{Q}[c][c]{$\langle \Psi| $}
\psfrag{O}[c][c]{$O$}
  \includegraphics[width=.95\columnwidth]{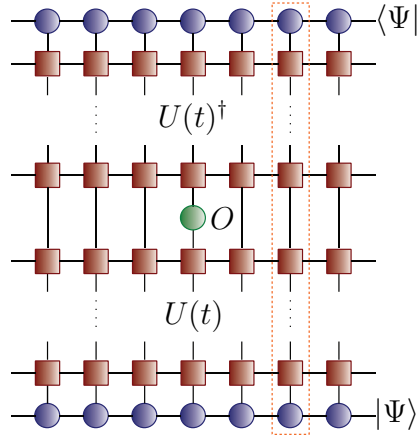}
}
\end{minipage}
\caption{Construction of the two-dimensional tensor network representing a time-dependent expectation value.}
\label{fig:fig2}
\end{figure}

By applying the TN approximated evolution operator to the initial
state, $|\Psi(0)\rangle$, we obtain a TNS, belonging to the class of concatenated tensor
states (CTN) introduced in~\cite{huebener09ctn}, which describes the evolved
state [\fref{fig:fig2-a}]. The description is exact up to the error introduced by the
Trotter decomposition.
Higher order
Suzuki-Trotter approximations are also
possible~\cite{suzuki90,suzuki91oT3po} which 
reduce the error for a given step, $\delta$, by involving the product of a larger
number of exponentials per step.

We can then construct the tensor network that represents the expectation value of any time-dependent observable,
$\langle O(t)\rangle=\langle\Psi(t)|O|\Psi(t)\rangle$, by applying
the operator $\hat{O}$ to the TNS for the evolved state and then contracting
with its adjoint. Graphically, we
represent such contraction in \fref{fig:fig2-b}.

The time evolution created by any short-range Hamiltonian acting on an
initial MPS will have a description in this form.  Although different
decompositions of the Hamiltonian will lead to a different expression for
the operators $U(\delta)$, as long as a MPO approximation is found for
them, the scheme described here will still be valid.
Also imaginary time evolution accepts an analogous tensor network representation in terms of
a succession of MPO, so that a similar tensor network can also
describe equilibrium observables, corresponding to ground or thermal
states, as we will discuss in more detail in the following sections.
The construction can also account for the
evolution under a time-dependent Hamiltonian.  In that case the evolution operator is
given by a time-ordered exponential, and can also be approximated by discrete steps, to which 
corresponding MPO decompositions can be applied that 
approximate the evolution operators to the desired order~\cite{suzuki93oT4}.

\subsection{\label{sec:2Dcontraction}Contracting the 2D tensor network}

The above discussion reduces the problem of computing the expectation
value $\langle O(t)\rangle$ to that of contracting the 2D tensor
network in \fref{fig:fig2-b}.  Contracting a 2D arbitrary tensor
network is known to be in general an extremely hard problem, in the
$\#P$-complete complexity class~\cite{schuch07comput}.  However, for
particular instances there may exist a contraction strategy which
allows us to approximate the result or even, in some cases, to compute
it exactly, where other strategies seem to fail.

In general, any strategy for the contraction of a 2D tensor network
could be used to approximate the expectation value $\langle
O(t)\rangle$. The degree of success of a given approach will depend
on the problem and, in particular, on how well the chosen algorithm
can take account of the entanglement structures that can be identified
in the network.

The contraction of a 2D tensor network appears as a fundamental routine in
the numerical simulation of 2D systems using TNS.
 It appears in the computation of classical partition
functions using DMRG based methods~\footnote{Actually, the structure
  of the tensor network generated by imaginary time evolution is
  analogous to that discussed in the context of transfer-matrix
  DMRG~\cite{bursill96trans}, while the class of 2D tensor networks
  appearing for instance in the contraction of PEPS can have a more
  complex structure.}, where approaches like transfer matrix DMRG~\cite{wang97trans,bursill96trans} and
corner transfer matrix~\cite{nishino96ctm} techniques have been developed.
In the context of the numerical simulation of quantum systems, different algorithms have been proposed in the last years~\cite{verstraete04peps,zhao10trg,wang11MCpeps,orus09ctm,garcia113d}.

\begin{figure}[floatfix]
\subfigure[Standard contraction strategy along the time direction.]{
 \label{fig:fig3-a}
  \includegraphics[width=.7\columnwidth]{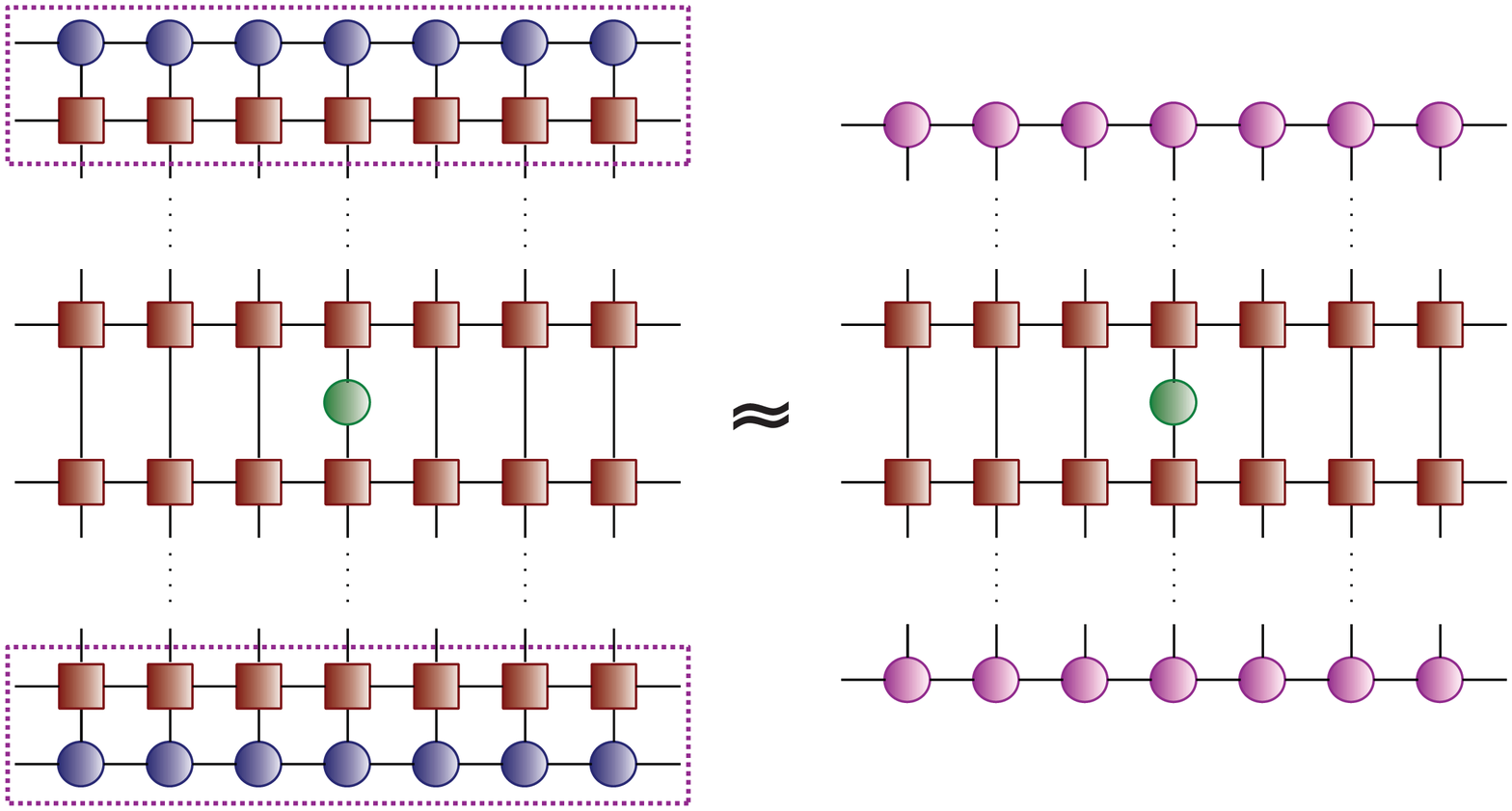}
}
\subfigure[Contraction in the Heisenberg picture.]{
 \label{fig:fig3-b}
  \includegraphics[width=.7\columnwidth]{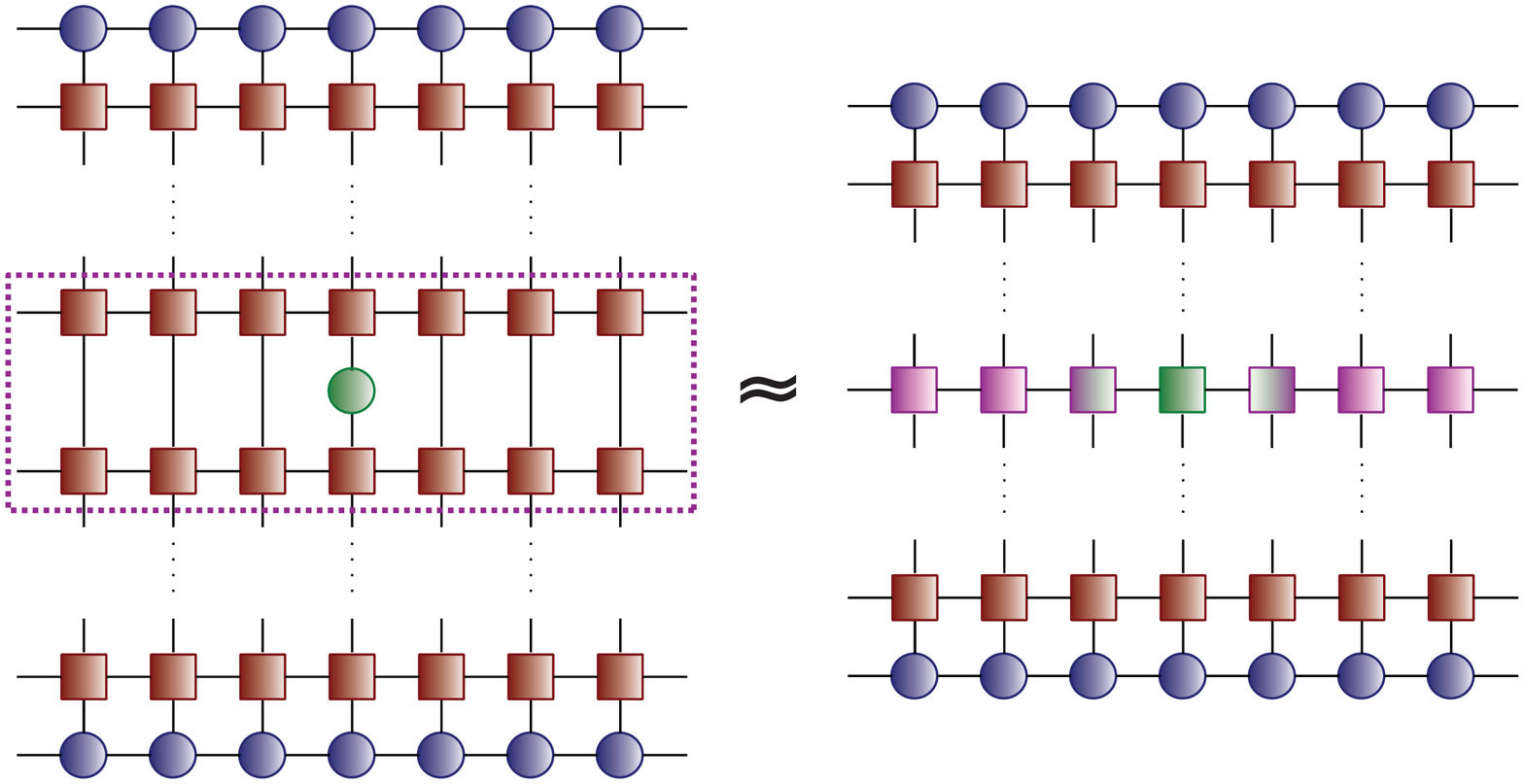}
}
\caption{Standard (tDMRG) and Heisenberg picture strategies to
  contract the tensor network}
\label{fig:fig3}
\end{figure}

The currently standard strategy for the study of time dependent
quantum many-body systems consists in approximating the evolved state,
after every time step, by a new MPS. This is the basic building block of TEBD and 
time-dependent DMRG (tDMRG) algorithms~\cite{vidal03eff,daley04adapt,white04realt},
and is indeed a particular way of performing the contraction, namely in the time direction.  As
illustrated in \fref{fig:fig3-a}, by approximating the 
product of one evolution step times the state by a new MPS, the height of the TN is reduced.
At any time along the evolution, the expectation value can be computed by
contracting a local operator between the evolved MPS and its adjoint.
However, the entropy in the evolved state may grow as fast as linearly
with time for out-of-equilibrium dynamics~\cite{calabrese05,schuch08entropy}. 
Since the entanglement encompassed by a MPS ansatz is bounded by the logarithm
of the bond dimension, $\log D$, and thus the computational effort required
to accurately describe the evolved state as a MPS will grow exponentially, making it impossible for the
algorithm to follow the evolution beyond short times. In practice,
this is evident by an abrupt onset of the truncation
error~\cite{schollwoeck06tDMRG}.

A different contraction strategy is the evolution of local operators using 
tDMRG in the Heisenberg picture~\cite{clark10heis,muth11heis}. As
illustrated in \fref{fig:fig3-b}, this corresponds to contracting
the problem tensor network also in the time direction, but from inside
out, successively approximating the evolved operator, $\hat{O}(t)$, by
a MPO.  This strategy can beat the standard
tDMRG approach when the evolved operators have a good approximation, or
even an exact expression, as MPO, while the evolved state shows a
linear increase of entanglement~\cite{prosen07ising,pizorn09xy}, but
in more general cases may require also exponential resources~\cite{prosen07integ}.

Other time evolution methods have been proposed based on a light cone 
strategy~\cite{hastings08lightcone,enss12lightcone}. They can also be understood as the 
contraction of an effective 2D tensor network.

\begin{figure}[floatfix]
\begin{minipage}[b]{.35\columnwidth}
\subfigure[Transverse contraction along space direction.]{
 \label{fig:fig4-a}
\psfrag{contractL}[c][c]{contraction}
\psfrag{contractR}[c][c]{}
\psfrag{O}[c][]{$E_O$}
\psfrag{E}[c][]{$E$}
\psfrag{L}[c][]{$\langle L |$}
\psfrag{R}[c][]{$|R\rangle$}
  \includegraphics[height=.9\columnwidth]{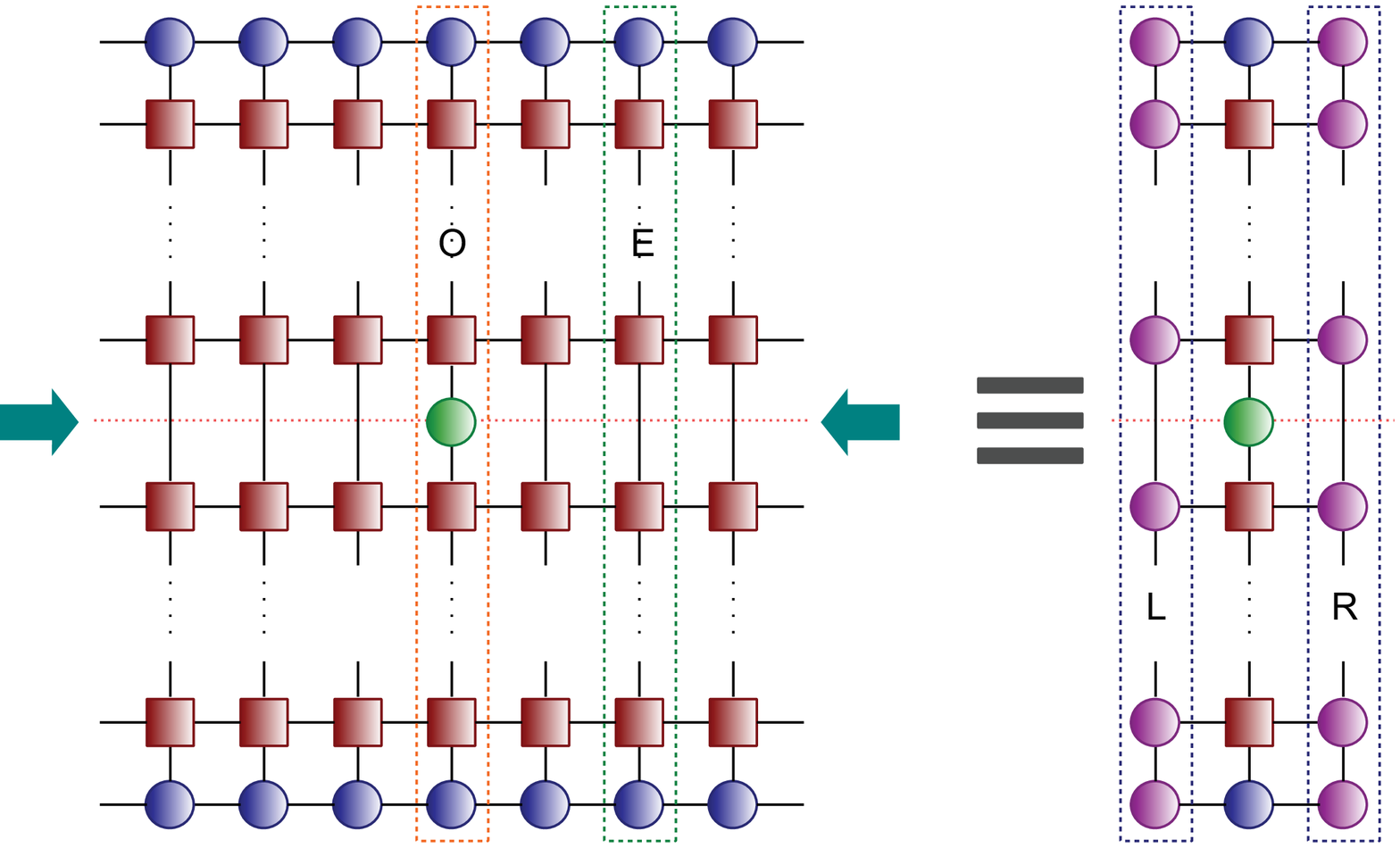}
} 
\end{minipage}
\hspace{.25\columnwidth}
\begin{minipage}[b]{.3\columnwidth}
\subfigure[Folding of the network.]{
 \label{fig:fig4-b}
\psfrag{O}[bc][bc]{$\tilde{E}_O$}
\psfrag{E}[bc][bc]{$\tilde{E}$}
\psfrag{L}[c][c]{$\langle \tilde{L} |$}
\psfrag{R}[c][c]{$|\tilde{R}\rangle$}
\psfrag{F}[cl][cl]{\parbox{.2\columnwidth}{folding axis}}
  \includegraphics[height=.9\columnwidth]{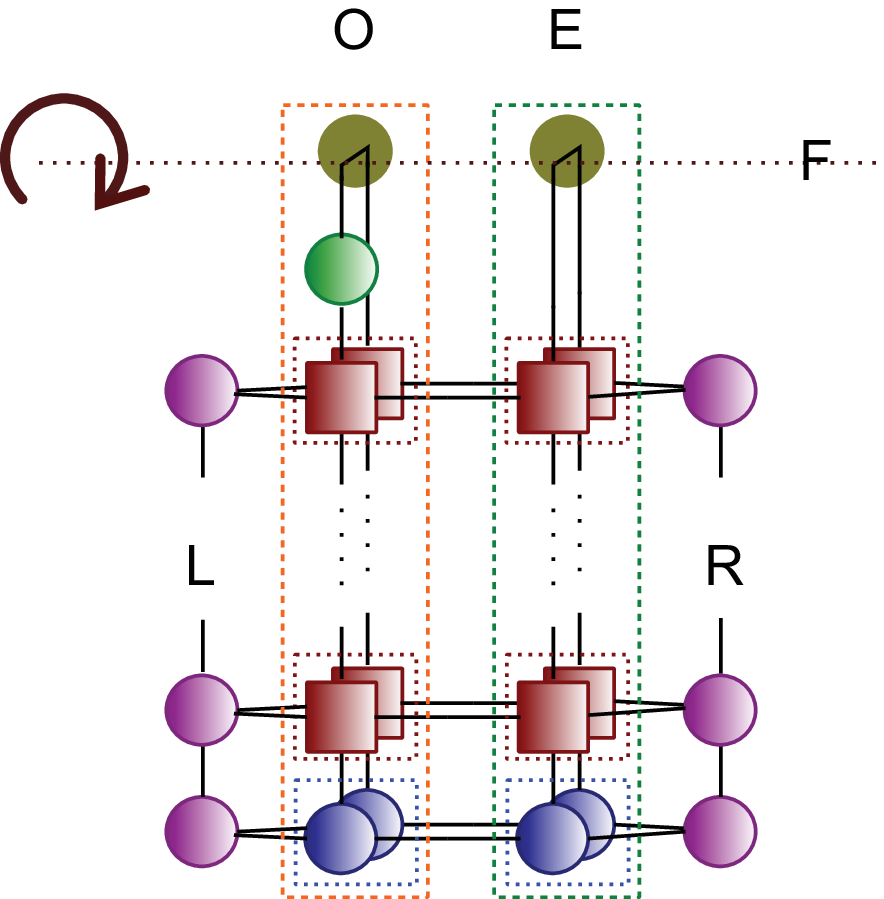}
}
\end{minipage}
\caption{Contraction of the network for a local expectation value
  $\langle O(t)\rangle$ with the basic transverse strategy (a) and 
  folding (b), where operators for the same time step are grouped
  together in a double effective operator.  }
\label{fig:fig4}
\end{figure}

Here we will focus on the transverse folding technique, introduced
in~\cite{banuls09fold}.  Different to the approaches mentioned above,
in this algorithm the tensor network is contracted along the spatial
direction [see \fref{fig:fig4}], thus avoiding the explicit
truncation of the evolved state. In this scheme no TNS
approximation for the state $|\Psi(t)\rangle$ is explicitly built.

Let us assume that the operator $\hat{O}$ is a single-body observable acting
on site $M$.~\footnote{The argument can be immediately generalized to
  products of local observables.}  By construction, every
column in the tensor network is a MPO. For any site $k\neq M$, this
MPO exactly corresponds to the transfer matrix of the evolved
state~\cite{perez07mps}, $E^{[k]}(t)=\sum_i {\bar{A}}^{[k]^i}(t)\otimes
A^{[k]^i}(t)$, where $A^{[k]^i}(t)$ has itself a MPO structure, or more precisely, is a CTN 
as defined in~\cite{huebener09ctn}.
Indeed, $A^{[k]^i}(t)$ is
given by the local operators that act on site $k$ at all times, and represents exactly 
the evolved MPS.  For site $M$, the MPO includes the application
of the operator, $E_O^{[M]}(t)=\sum_{i,j}[{\bar{A}}^{[M]^i}(t)\otimes
  A^{[M]^j(t)}] \langle i|O|j\rangle$.  In terms of the transfer
matrices, the expectation value can be written 
\beq 
\langle \Psi(t)|O|\Psi(t)\rangle =
\mathrm{tr}(E^{[1]} 
\ldots E_O^{[M]} \ldots
E^{[N]}). 
\label{eq:transO}
\eeq

In the case of a finite chain, the contraction can be performed in an approximate way using
the technique introduced
in~\cite{verstraete04peps,murg07hard}.
The leftmost term, $E^{[1]}$, is a
vector, with MPS structure.  The product $E^{[1]}E^{[2]}$ is then read as the
left action of a MPO on a MPS, which can be approximated by a MPS.
Iterating this and proceeding in the same way for the right part of the network, 
$\langle \Psi(t)|O|\Psi(t)\rangle = \langle L_M|E_O^{[M]}|R_M\rangle$,
where $\langle L_M|$ and $|R_M\rangle$ represent the vectors resulting
from the contraction of the half-networks to the left and right of
site $M$, respectively [see \fref{fig:fig4-a}]. If these vectors are approximated by MPS, the
remaining contraction can be done exactly in an efficient way.

\subsection{\label{sec:transverse}Transverse contraction for an infinite chain}

The transverse approach is particularly convenient when dealing
with an infinite chain. In this case, the resulting 2D tensor network
for a given total evolution time $t$ is infinite in space, but finite
in the time direction.  Contraction techniques that follow the time
direction typically rely on the translational invariance of the
system~\cite{ostlund95td,vidal07infinite} for this scenario.  This
limits their applicability in situations that break the symmetry, as in the presence of impurities.  The efficiency of the
contraction in the space direction, in contrast, is not affected if
the breaking of the translational symmetry occurs 
only for a finite sub-chain. Cases where the chain is periodic can also be treated in 
a similar manner.

In the thermodynamic limit, the time-dependent observable \eqref{eq:transO} will be given by 
$$ \langle \Psi(t)|O|\Psi(t)\rangle = \lim_{n\rightarrow\infty}
\mathrm{tr}(E^{[-n+1]} \ldots E_O^{[M]} \ldots E^{[n-1]}). $$ If the
system is translationally invariant, all
columns will be identical, $E^{[k]}\equiv E(t)$, except for the one containing the application
of the operator, $E_O$.  If $\lambda$ is the eigenvalue of the
transfer matrix $E$ with the largest magnitude, and it is the only one with this absolute value (the generic case
for MPS~\cite{perez07mps}), $ \left [E(t)\right ]^k \sim \lambda^k|R\rangle\langle L| $ for large $k$.  The infinite right and
left half networks can thus be reduced to vectors which are
proportional to the dominant right and left eigenvectors of $E(t)$,
respectively $|R\rangle$ and $\langle L |$. The proportionality factors 
can be eliminated by normalizing,
\beq
\langle O(t) \rangle=\frac{\langle \Psi(t)|O|\Psi(t)\rangle}{\langle \Psi(t)|\Psi(t)\rangle}
=\frac{\langle L | E_O | R \rangle}{\langle L | E | R \rangle}.
\label{eq:O(t)}
\eeq 
The networks that represent the numerator and denominator differ only in the 
single site where the operator acts, which for the latter contains the identity in the place of $\hat{O}$.
If an even-odd decomposition of the Hamiltonian is used, the network is not translationally invariant, but we can recover this property blocking together pairs of sites, and the above procedure is still valid.

The dimension of the vector space on which operators $E(t)$  act
grows exponentially with the number of time steps, so that in general it
will not be possible to compute $|R\rangle$ and $\langle L |$ exactly.
However, finite MPS tools can be used also in this case to perform the contraction in
the space direction and find MPS approximations to the dominant
eigenvectors.  In particular, we implement the power method, by repeatedly
applying the transfer matrix $E(t)$ (already written as a MPO along
the time direction, see \fref{fig:fig4-a}) to the left and to the
right of an arbitrary initial MPS vector and truncating the result to a
given bond dimension, $D$, as described
in~\cite{verstraete04mpdo,murg07hard}, until convergence is attained.
This procedure yields a MPS approximation to the eigenvectors, with the
truncation taking always place in the space of transverse vectors.
The result of \eqref{eq:O(t)} is then obtained by computing two
expectation values of MPO in states given by MPS.

If the non-degenerate dominant eigenvector of the transfer matrix
$E(t)$ can be well approximated by a MPS with small bond dimension, the
procedure described above will efficiently yield a good
approximation to the time-dependent observable. But if the bond
dimension required for the MPS approximation grows fast with time, the
method will face a similar problem to the standard contraction in the
time direction.
Since, roughly speaking, MPS are best at describing states with small entanglement, the 
success of this approach will then be related to the amount of entanglement in the transverse dominant eigenvectors.
We will discuss next how, in some cases, this amount can be dramatically reduced by folding.

\subsection{\label{sec:folding}Folding the tensor network}

The folding technique combines the transverse contraction 
described above with a more efficient representation of the
entanglement in the dominant eigenvectors.  
This strategy can be physically motivated by the picture of a freely
propagating excitation, initially localized at one site of the chain.
Although the initial state is a product, in the evolved state after
time $t$, sites at a distance proportional to $2 t$ become entangled.
In the transverse direction this translates, after contracting from the right until a given position, $x$,
into entanglement between time sites 
that correspond
to the instant in which the excitation reaches $x$.
Folding the network in two along the space-like line for the final time [see \fref{fig:fig1-b}]
groups these sites together, and thus removes
all the transverse entanglement. 

On a given MPO 
column in \fref{fig:fig4-a}, tensors that represent the same factor
of the unitary decomposition, for the same time step in the state and
its adjoint, are located at the same distance of the center of the
network (corresponding to the final time, $t$). The folding
transformation brings together these pairs and defines a new MPO, as
shown in \fref{fig:fig4-b}.

The folding operation can also be explained as the equivalent
contraction $ \langle \Psi(t)|O|\Psi(t)\rangle=
\langle\Phi | O\otimes \ident \left ( |\Psi(t)\rangle\otimes|\bar{\Psi}(t)\rangle\right), $
where $|\bar{\Psi}(t)\rangle$ is the complex conjugate of
$|\Psi(t)\rangle$.  
 The bra corresponds to the product of (unnormalized) maximally entangled pairs
between each site of the chain and its conjugate,
$|\Phi\rangle=\left[\sum_{i_k=1}^d|i_k\bar{i}_k\rangle\right]^{\otimes N}$.  We can
now group together each tensor in the $|\Psi(t)\rangle$ network  
 with the corresponding one in the adjoint, and define the effective tensors of the folded network in which the
bond and physical dimensions are squared.

After folding, the network can be contracted using the transverse technique
above. If the pattern of entanglement is similar to that in the intuitive picture, 
it will be possible to find good approximations to the dominant eigenvectors
using MPS of reduced bond dimension.

\section{\label{sec:entanglement}Understanding the entanglement in the network}

As evident from the discussion above, the various MPS techniques to compute time dependent expectation values will 
be sensitive to the amount of entanglement in the vectors or operators that each one of them needs to approximate. Here we will focus on the outside-in approaches in the time and space directions, where the partial contraction of the TN involves MPS approximations to different vectors. In the Heisenberg picture approach, the relevant quantity is the entanglement in the operator space, discussed in~\cite{prosen07ising}.

To better understand the different entanglement structures that can appear in the evolution network when contracted along different directions, we introduce a simple toy model, which mimics the
dynamics of freely propagating excitations via a tensor network. We analyze the entanglement to which this simple
model gives rise in each contraction direction.  Although a
true Hamiltonian dynamics generates a more complicated tensor network,
this picture is useful to give us some intuition about the
entanglement patterns that time-like and space-like strategies can
handle efficiently.  It also shows how identifying the main
entanglement contribution may be the key to an efficient accurate
approximation of the result of the contraction.


In the second part of the section, we make this statements more
quantitative by looking at the real time evolution of a spin chain
under a family of nearest-neighbour Hamiltonians.  We consider in
particular, a family of Hamiltonians \beq H=-\sum_i\left(\sigma_z^i
\sigma_z^{i+1}+g \sigma_x^i+h \sigma_z^i \right ),
\label{eq:Ising}
\eeq with Ising $ZZ$ interactions and a magnetic field which may have
components in the transverse, $g$, and parallel, $h$, directions.
This family has two well studied limits.  When $h=0$, we recover the
integrable Ising model in a transverse field, critical at $g=1$, and
for which the exact solution is known, whereas for $g=0$ we obtain a
classical model, as the Hamiltonian is diagonal in the computational
basis.  Intermediate values of $g$ correspond to non-integrable
models.

\subsection{\label{sec:toy}A tensor network toy model}

\begin{figure}
\psfrag{L}[c][c]{$A_L$}
\psfrag{R}[c][c]{$A_R$}
\psfrag{a}[cl][cl]{$\alpha_{\ell}$}
\psfrag{b}[cl][cl]{$\beta_{\ell}$}
\psfrag{c}[cl][cl]{$i_{\ell}$}
\psfrag{d}[cl][cl]{$j_{\ell}$}
\psfrag{e}[cl][cl]{$\alpha_{r}$}
\psfrag{f}[cl][cl]{$\beta_{r}$}
\psfrag{g}[cl][cl]{$i_{r}$}
\psfrag{h}[cl][cl]{$j_{r}$}
\psfrag{A}[l][l]{$\equiv \tilde{A}_{\tilde{\alpha}\tilde{\beta}}^{\tilde{i}\tilde{j}}$}
  \includegraphics[width=.45\columnwidth]{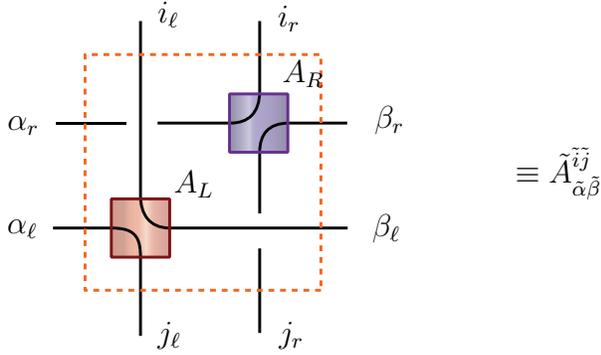}
\caption{Simple MPO that mimics the dynamics of freely propagating excitations.}
\label{fig:toympo}
\end{figure}

We consider sites with physical dimension $d=4$. We can represent each
site as composed by two two-level systems, $|a\rangle_{\ell}\otimes |b
\rangle_{r}$, and identify the \emph{occupied} state of the left
(right) subsystem, $|1\rangle_{\ell(r)}$, with the presence in this
position of a particle propagating to the left (right).  The dynamics
of these particles can be modelled by a simple MPO which acts
as a tensor product on left- and right-moving particles, as depicted in
\fref{fig:toympo}, 
\beq
\tilde{A}_{\tilde{\alpha}\tilde{\beta}}^{\tilde{i}\tilde{j}} \equiv
\left( A_L \right)_{\alpha_{\ell}\beta_{\ell}}^{i_{\ell} j_{\ell}}\otimes
\left( A_R \right)_{\alpha_{r}\beta_{r}}^{i_{r} j_{r}},
\label{eq:toympo}
\eeq
where $\tilde{i}\equiv i_{\ell}i_r$, and all the indices of tensors $A_{L}$ and $A_R$ have dimension two. The 
only non-vanishing components or the tensor are
\bea
\left( A_L \right)_{00}^{00}=\left( A_L \right)_{10}^{10}=
\left( A_L \right)_{01}^{01}=\left( A_L \right)_{11}^{11}&=&1, \nn\\
\left( A_R \right)_{00}^{00}=\left( A_R \right)_{10}^{01}=
\left( A_R \right)_{01}^{10}=\left( A_R \right)_{11}^{11}&=&1.
\nn 
\eea

This MPO is a unitary transformation which simply shifts type $\ell$
particles one site to the left, and type $r$ ones one site to the
right. The graphical representation (\fref{fig:toympo}) 
encompasses the actual action of the tensors, which swap the state of the physical (vertical) indices to the virtual (horizontal) ones. 
It is possible to easily characterize how entanglement
develops in this tensor network depending on the initial state. 

First we consider an initial product state in which each site is in a
maximally entangled state of one particle moving to the left and one
moving to the right,
$\frac{1}{\sqrt{2}}\left(|1\rangle_{\ell}|0\rangle_{r}+|0\rangle_{\ell}|1\rangle_{r}\right)$.
This mimics the intuitive picture of initially localized excitations
that create entanglement when travelling away.  Because the evolution
affects left- and right-particles independently, we immediately see 
that after $t$ applications of the MPO the maximally entangled
pairs, which were initially localized, stretch over a distance $2
t$. Therefore the entanglement across any cut of the chain will grow
linearly with time, and the bond dimension of the evolved state will
grow exponentially.

\begin{figure}[floatfix]
\subfigure[]{
\label{fig:fig5-a}
\begin{minipage}[c]{.3\columnwidth}
\psfrag{k}[c][c]{$k$}
\psfrag{b}[c][c]{$\bar{k}$}
  \includegraphics[height=1.5\columnwidth]{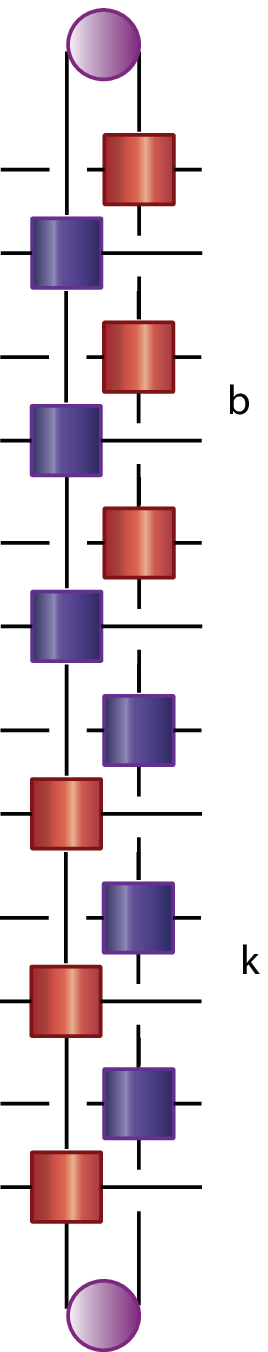}
\end{minipage}
}
\subfigure[]{
\label{fig:fig5-b}
\begin{minipage}[c]{.3\columnwidth}
\psfrag{k}[c][c]{$k$}
\psfrag{l}[c][c]{$\ell_k$}
\psfrag{r}[c][c]{$r_k$}
  \includegraphics[height=1.5\columnwidth]{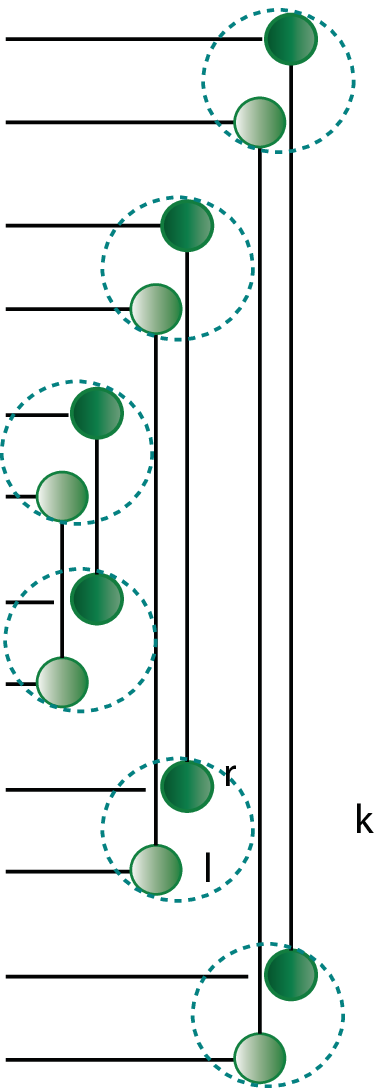}
\end{minipage}
}
\subfigure[]{
\label{fig:fig5-c}
\begin{minipage}[c]{.3\columnwidth}
\psfrag{k}[c][c]{$k$}
\psfrag{l}[c][c]{$\ell_k$}
\psfrag{r}[c][c]{$r_k$}
  \includegraphics[height=1.5\columnwidth]{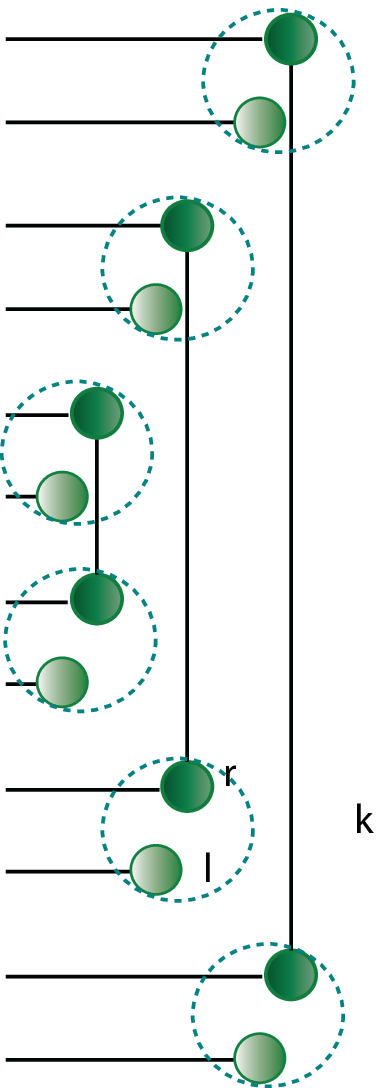}
\end{minipage}
}
\caption{Entanglement structures in the toy TN model. In (a) we show the transverse matrix $E(t=3)$ for the toy model tensor network~\eqref{eq:toyE}; in (b) and (c) we represent the dominant right eigenvectors for this operator in the cases, respectively, of a maximally entangled and a product initial state, as described in the text.}
\label{fig:fig5}
\end{figure}

We can easily compute the transverse dominant eigenvectors of the evolution
network. The tensors acting on the transverse vectors are obtained
by rotating $A_L$ and $A_R$ 90 degrees to the right, which has the effect of 
exchanging $L$ and $R$ terms. Taking into account that the adjoint of the MPO is also obtained by 
exchanging both terms, the transfer matrix will have a structure [\fref{fig:fig5-a}]
\beq
E(t)=\langle B |  \left(A_R \stackrel{(t)}{\ldots} A_R A_L \stackrel{(t)}{\ldots} A_L\right)
\otimes\left(A_L \stackrel{(t)}{\ldots} A_L  A_R \stackrel{(t)}{\ldots} A_R \right) |\bar{B}\rangle ,
\label{eq:toyE}
\eeq
where $B$ is the tensor corresponding to the initial state 
 and we have omitted the explicit left and right indices of each tensor for simplicity. 
 The tensor product of the middle term reflects the separable evolution of the left-
and right-particles. Both components can only be connected by the
initial state which, in the case we are considering, is the
maximally entangled state between both, so that the tensor $B$ has components $(B_{\Phi})_{i_{\ell} i_r}=\frac{1}{\sqrt{2}}\delta_{i_{\ell}i_r}$.  It is easy to check that the
following state is invariant under the transfer matrix, so it is an eigenvector with unit 
eigenvalue,
\beq
|R_{\Phi} \rangle = \otimes_{k=1}^t  \left [ \frac{1}{\sqrt{2}}\left ( |0\rangle_{\ell_k}|0\rangle_{\ell_{\bar{k}}} +
|1\rangle_{\ell_k}|1\rangle_{\ell_{\bar{k}}} \right )  \otimes \frac{1}{\sqrt{2}}
\left ( |0\rangle_{r_k}|0\rangle_{r_{\bar{k}}} +
|1\rangle_{r_k}|1\rangle_{r_{\bar{k}}} \right )\right],
\eeq
where $|0\rangle_{\alpha_k}$ represents the state of the $\alpha$
($\ell$ or $r$) subsystem at the time site corresponding to the $k-th$
application of the MPO.  The subindex $\bar{k}$ refers to the
symmetric time site, i.e. the $k-th$ application of the adjoint MPO.

Then, for each time step $k$, the dominant eigenvector contains a
product of two maximally entangled pairs, one in the left and one in
the right component of the network, stretching over a distance
$2(t-k)$, which are represented in \fref{fig:fig5-b}. The bond
dimension required to write this vector as a MPS will then grow
exponentially with time.  

After folding, time sites $k$ and
$\bar{k}$ are grouped in an effective site $\tilde{k}$, which will
have folded $\tilde{\ell}$ and $\tilde{r}$ components.  The dominant
eigenvector in the folded representation can be written as 
\beq
|R_{\Phi}\rangle=\otimes_{k=1}^t \left [ |\Phi_+ \rangle_{\tilde{\ell}_k}\otimes |\Phi_+
\rangle_{\tilde{r}_k} \right ] , 
\eeq 
where $|\Phi_+
\rangle_{\tilde{\ell}_{k}}=\frac{1}{\sqrt{2}}\left (|0\rangle_{\ell_k}|0\rangle_{\ell_{\bar{k}}}
+ |1\rangle_{\ell_k}|1\rangle_{\ell_{\bar{k}}} \right )$ is a pure
state of the $\ell$ component of the new effective site $\tilde{k}$,
which contains exactly the two maximally entangled left-type sub-sites of $k$ and $\bar{k}$,
so that the eigenvector is written as a product state.

It is also interesting to consider the opposite limit, a case in which
the entanglement will not increase with time. In particular, we have
considered the product state $|1\rangle_{\ell} |1\rangle_r$.  This is
an eigenstate, invariant under the evolution MPO, and contains no
entanglement in space.  Nevertheless, the dominant eigenvector of the
transfer matrix can be checked to be of the form [see
\fref{fig:fig5-c}]
\beq 
|R_{1\otimes 1}\rangle=
\otimes_{k=1}^t \left [  |1\rangle_{{\ell}_k}
\otimes |1\rangle_{{\ell}_{\bar{k}}} \otimes \frac{1}{\sqrt{2}}\left (
|0\rangle_{r_k}|0\rangle_{r_{\bar{k}}} +
|1\rangle_{r_k}|1\rangle_{r_{\bar{k}}} \right )\right ] .  
\eeq Therefore the
transverse vector contains also nested maximally entangled pairs, in a
number that grows linearly with $t$, but only half the
maximum possible amount, which was attained in the case before.  In a
situation like this, the transverse contraction of the original
network will fail, while the standard contraction is trivial.
However, folding the network reduces again the dominant eigenvector to
a product, thus allowing the transverse contraction.  The initial
product $|0\rangle_{\ell} |0\rangle_r$, which could be understood as
having no excitation in the system, behaves in an analogous way.

\subsection{\label{sec:entropy}Entropy in the transverse contraction for Ising-like Hamiltonians}

To make our statements more quantitative, we have compared the
behaviour of entanglement in the transverse contractions, with and
without folding, for time evolution under the family of Hamiltonians
defined in \eqref{eq:Ising}.  As a figure of merit we compute the
maximum entanglement entropy, $S_{\mathrm{max}}$, with respect to all
possible bipartitions, in the dominant right
eigenvectors~\footnote{Right and left eigenvectors show similar
  behaviour.} at different times, and we compare this quantity before
and after the folding for different Hamiltonians and initial states.
Although the truncation error of the dominant eigenvectors
would properly evaluate the accuracy of the transverse approaches, this 
entropy has a more physical interpretation. It is the 
analogue to the entanglement in the time direction, and allows us
to connect our observations to the intuitive models. 

In particular, we have studied three representative scenarios. First, we simulate a
sudden quench in the integrable Ising model, starting from an initial
product state, for which the entropy in the standard picture is known
to grow linearly with time.  Second, we take the opposite limit, and
set the initial state to be the ground state of the evolving
Hamiltonian.  Naively, these two scenarios correspond to the cases
described for the toy model in the previous section.  Finally, we
consider a more general quench, in which the initial state is the
entangled ground state of \eqref{eq:Ising} for certain values of $g$,
$h$, and evolves under a different set of parameters, not
corresponding to any integrable limit.

\begin{figure}[floatfix]
\resizebox{\columnwidth}{!}{\includegraphics{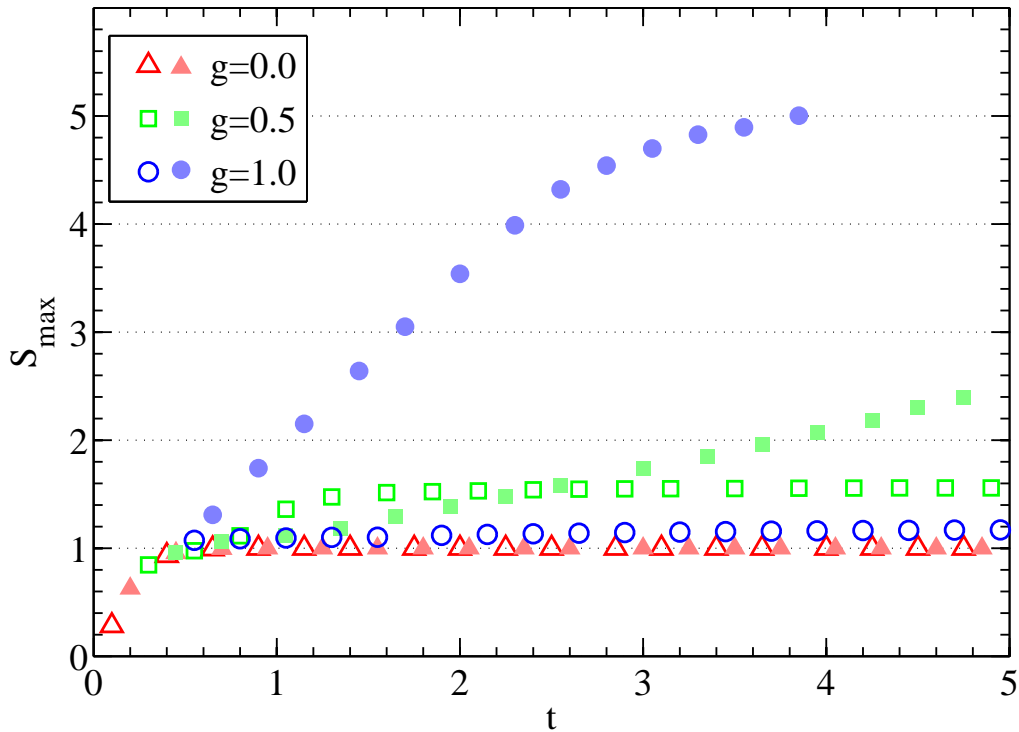}}
\caption{ Maximal entropy in the transverse dominant eigenvectors as a
  function of time in the evolution of a product state $|X+\rangle$
  under the exactly solvable Ising model (vanishing parallel field
  $h=0$) with transverse field $g=1$ (blue circles), $0.5$ (green squares) and $0$ (red triangles). The solid symbols
  correspond to the original network and the empty ones to the folded
  case.}
\label{fig:entropyProdg}
\label{fig:fig6}
\end{figure}

\begin{figure}[floatfix]
\resizebox{\columnwidth}{!}{\includegraphics{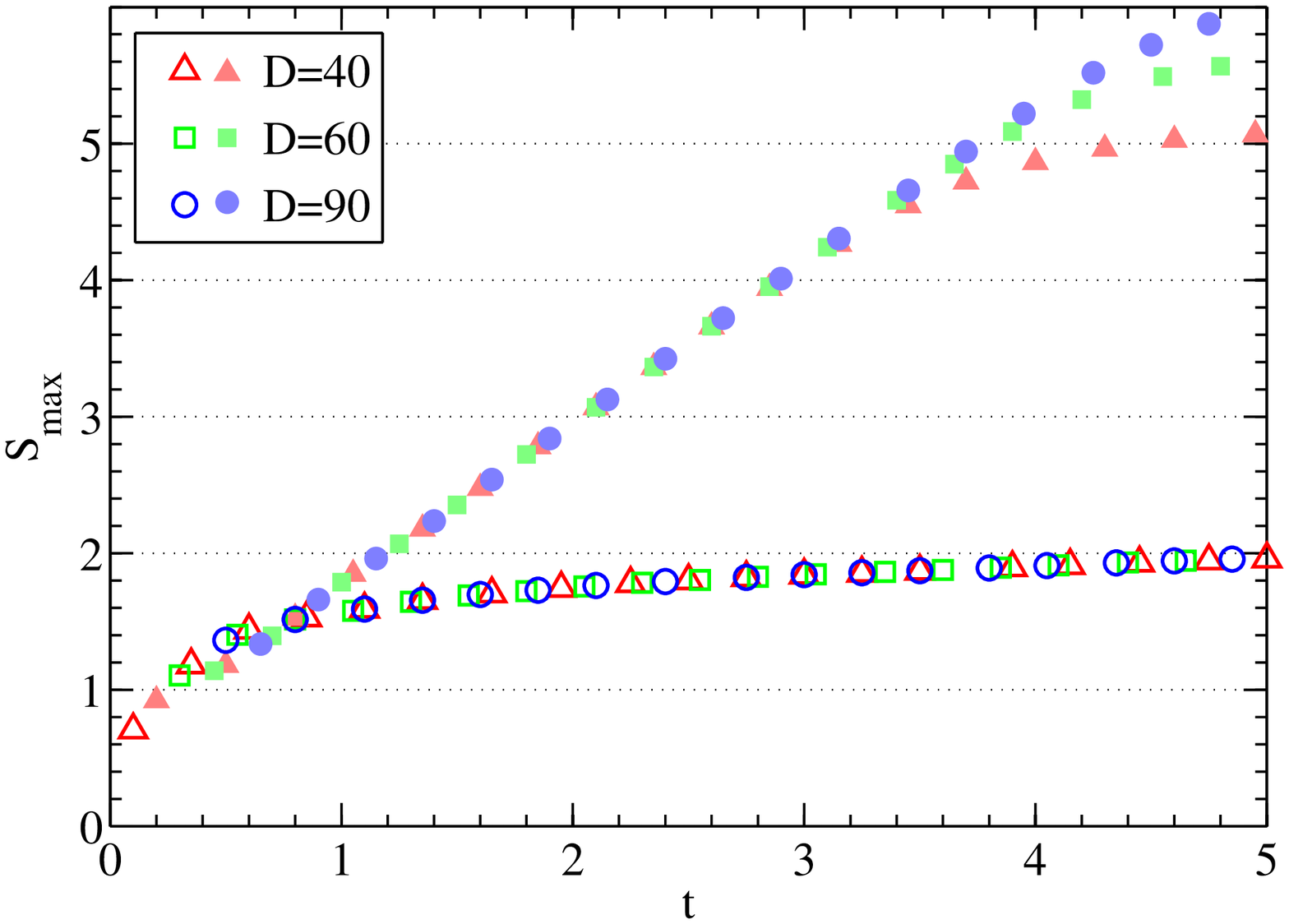}}
\caption{Maximal entropy in the transverse dominant eigenvectors as a function of time in
  the integrable Ising model with $g=1.1$, $h=0$, for initial ground
  state.  The solid (empty) symbols correspond to the original (folded) network, using bond dimension $D=40$ (red triangles), $60$
  (green squares) and $90$ (blue circles) for the eigenvector. 
  As apparent from the curves for the original network, a fixed $D$ can keep track of the linear increase in entropy only for a certain time, after which truncation is evident.}
\label{fig:entropyGSising}
\end{figure}

\Fref{fig:entropyProdg} shows the results corresponding to the first of these cases, when
the chain is initialized in a product state,
$|X+>\equiv\left[\frac{1}{\sqrt{2}}(|0\rangle+|1\rangle)\right]^{\otimes N}$, in
which all spins are aligned along the positive $\hat{x}$ direction, and evolves under the Ising
Hamiltonian~\eqref{eq:Ising} with fixed transverse field $g=1.1$ (and $h=0$).
In this limit, the Hamiltonian is equivalent to a free fermion model,
and the initial state $|X+\rangle$ corresponds to the fermionic
Fourier vacuum~\cite{sachdev}, containing a superposition of all
products of pairs of diagonal modes with opposite momenta $k$ and
$-k$. Therefore, one may expect that this scenario shows the closest
similarity to the intuitive picture of freely propagating excitations
described in \sref{sec:folding}. Actually, the entropy in
the evolved state is known to grow linearly with time. In our toy
model such a situation led to linearly growing entanglement in the
transverse contraction, which could be completely removed by folding.
From the numerical results we observe that the entropy in the
transverse eigenvector grows indeed linearly with time (the apparent
saturation obeys to truncation in the eigenvector at a given bond
dimension $D$, an effect shown explicitly in \fref{fig:entropyGSising}), and the maximal entropy of the folded eigenvector,
while having a value different from zero, saturates very fast, in
agreement with the intuitive expectation, and remains constant at
longer times.  The actual asymptotic value of
$S_{max}^{\mathrm{(folded)}}$ and the slope of
$S_{max}^{\mathrm{(unfolded)}}$ depend on the particular parameter $g$.
As $g$ decreases, the latter grows slower, until for $g=0$, the
classical model is recovered and both values coincide.

These observations corroborate the  simplified picture for
out-of-equilibrium situations in which we can expect excitations
propagating in the state and creating correlations at long distances
as time evolves.  In those scenarios, the growing correlation length
makes it difficult for a standard MPS approach to describe the evolved
state, and we thus expect the maximum gain from folding the tensor network.

The opposite situation would be that of an eigenstate of the
Hamiltonian, which along the evolution will only acquire a time-dependent phase.  If the
state was well described by an MPS, as will often be the case for the
ground state, it will still be a MPS at any later time, but the
transverse picture may introduce entanglement in the contraction, as
for the invariant state under the toy model in \sref{sec:toy}.  We have thus
looked at the entropy in the transverse eigenvectors, both with and
without folding, starting from an MPS approximation to the ground
state of the integrable Ising model ($g=1.1$, $h=0$).  The initial MPS
approximation was found via the iTEBD
algorithm~\cite{vidal07infinite} with bond dimension $\chi=8$, to an
accuracy in energy of the order or $10^{-7}$.  The entropy results are
shown in \fref{fig:entropyGSising}.  We observe good agreement
with the intuitive understanding from section~\ref{sec:toy}, with 
entropy growing linearly in the transverse contraction before folding,
whereas after folding it grows very slowly for long times, appearing
almost saturated.

Finally, we want to explore the entropy properties of a more general
case, which does not correspond to the toy model described above.  To
this end, we consider a more general dynamics ($h\neq 0$) where the
free fermion picture does not apply, and we simulate a global quench
in which we start from the ground state for $g=1.1$, $h=0$, as above,
and evolve the system under various Hamiltonians, with fixed $g$ and
varying $h$.

\begin{figure}[floatfix]
\resizebox{\columnwidth}{!}{\includegraphics{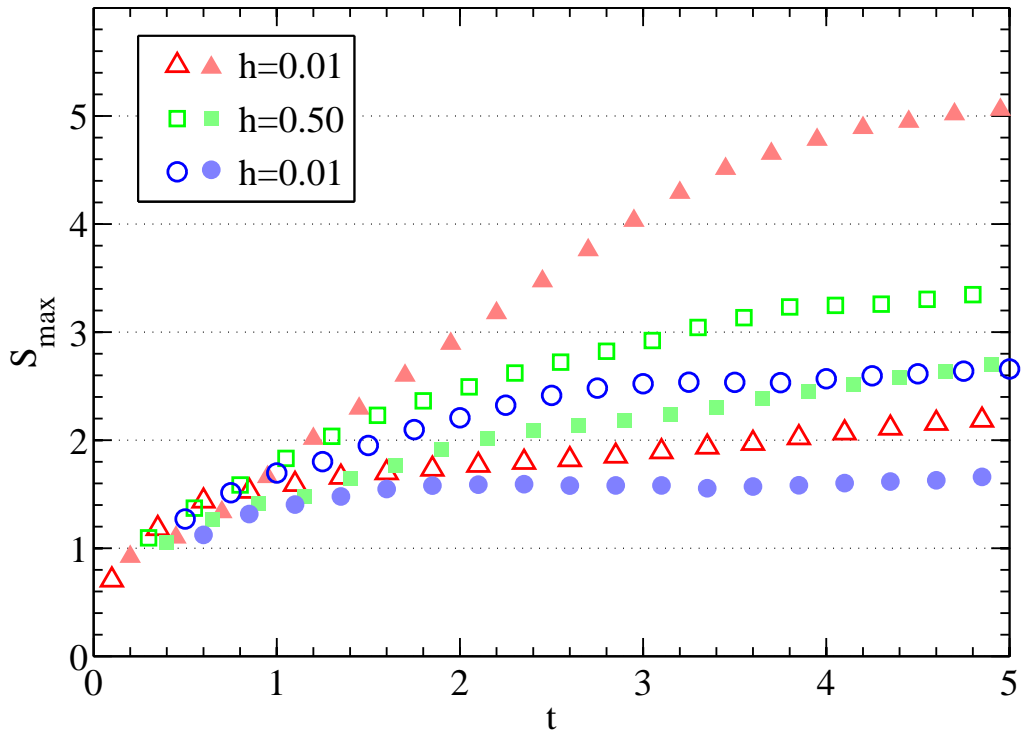}}
\caption{Maximal entropy in the transverse dominant eigenvectors as a
  function of time, after a quench of the parallel
  field $h$. The system starts in the ground state of
  \eqref{eq:Ising} with $g=1.1$, $h=0$ and evolves under a
  different Hamiltonian with fixed transverse $g=1.1$ and varying
  parallel field $h=0.01$ (red triangles), $0.5$ (green squares) and $1$
  (blue circles). Solid (empty) symbols correspond to the original (folded) network.}
\label{fig:entropyQuenchh}
\end{figure}

The results are shown in \fref{fig:entropyQuenchh}.  We observe
that for small $h$, the behaviour is close to that in
\fref{fig:entropyGSising}, most favourable for the folding
strategy.  As the parallel component of the magnetic field $h$ grows,
the difference between entropies before and after folding seems to
decrease, and at some point, the maximum entropy is smaller before the
folding. Even in these cases, the folded entropy lies well below the
upper bound $2 S_{\mathrm{max}}^{\mathrm{(unfolded)}}$, which would be
reached if there was no entanglement between the pairs of tensors the
folding brings together. A more determinant property than the absolute value of
$S_{\mathrm{max}}$ at a certain moment, at least concerning the
utility of these algorithms, is its scaling with time, as
this will ultimately determine the length of the evolution we can simulate
with these techniques.  Remarkably, for the case $h\simeq g$ we
observe that both curves have a similar pattern, with the entropy
saturating or growing very slowly for long times. This seems to
indicate that in this case, both techniques would support long time
calculations.

\section{\label{sec:applications}Applications and results}

In this section, we describe several physical problems in which the
transverse and folding techniques can be useful. 
 We adapt the techniques to the various scenarios and illustrate these applications with the
results for two families of spin models.

On one hand, we consider the Ising model, as defined
in \eqref{eq:Ising}. To complete our analysis, we present results also for 
simulations of the XY model, 
\beq 
H_{XY}=-\sum_i \left( J_x \sigma_x^{i} \sigma_x^{i+1}+ J_y
\sigma_y^{i} \sigma_y^{i+1} + J_z \sigma_z^{i}\right ), 
\label{eq:XY}
\eeq
 equivalent to a free-fermion case, too, and 
thus expected to show similar features to our simplistic model.

 In the following, we will exclusively consider chains in the thermodynamic limit.
 Although the basic techniques of folding and transverse contraction
 can be applied both to infinite and finite networks, the main
 advantage of using the folding strategy is due to the entanglement
 structure of the dominant eigenvectors that reduce the infinite
 network to a finite contraction.  In a finite case, on the contrary,
 the 2D network to be contracted spans the whole range of spatial
 sites, and in general the entanglement content of the transverse
 vectors will be complicated due to the boundary effects.

\subsection{\label{sec:outofeq}Out of equilibrium evolution}

A particularly interesting problem for these transverse strategies is
given by the out-of-equilibrium evolution of an infinite chain after a
global quench. This kind of dynamics can create fast growing
entanglement between spatial sites.  According to our intuitive model
and the entropy analysis discussed in the previous sections, the
transverse folding strategy should be best suited for this scenario,
being most advantageous when the scaling of the spatial entanglement
with time is the largest possible.

\begin{figure}[floatfix]
  \includegraphics[width=\columnwidth]{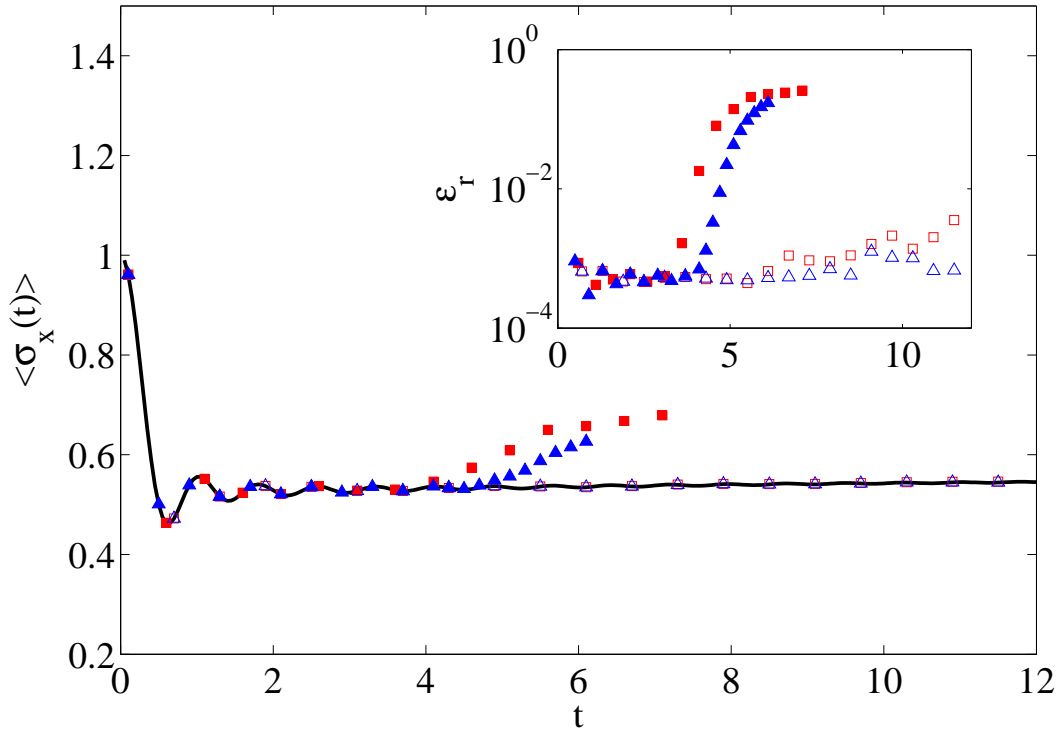}
\caption{Transverse magnetization per site $\langle \sigma_x(t)
  \rangle$ as a function of time for the evolution under the Ising
  model with $g=1.05$ of the initial product state $|X+\rangle$.  The
  solid line is the exact result, and the solid (empty) symbols show
  the results obtained with the transverse (folded) contraction using
  bond dimension $D=60$ (red squares) and $120$ (blue triangles) for
  the eigenvectors.  The inset shows the relative error.  }
\label{fig:polarizationIsing}
\end{figure}

Following from the study of the transverse entropies in the last section, 
a good realization of this intuitive picture occurs in the evolution,
under the exactly solvable Ising model, of an initial product state $|X+\rangle$. 
 In this situation, the Hamiltonian is
equivalent to a free-fermion model, and the entanglement in the space direction is known to
grow linearly with time. 
This scenario was already discussed
in~\cite{banuls09fold}, where we computed the time dependent local
magnetization as a function of time (shown in 
\fref{fig:polarizationIsing}),
 and checked
 the adequacy of the folding transverse
contraction to compute time-dependent local
observables.

\begin{figure}[floatfix]
\resizebox{\columnwidth}{!}{\includegraphics{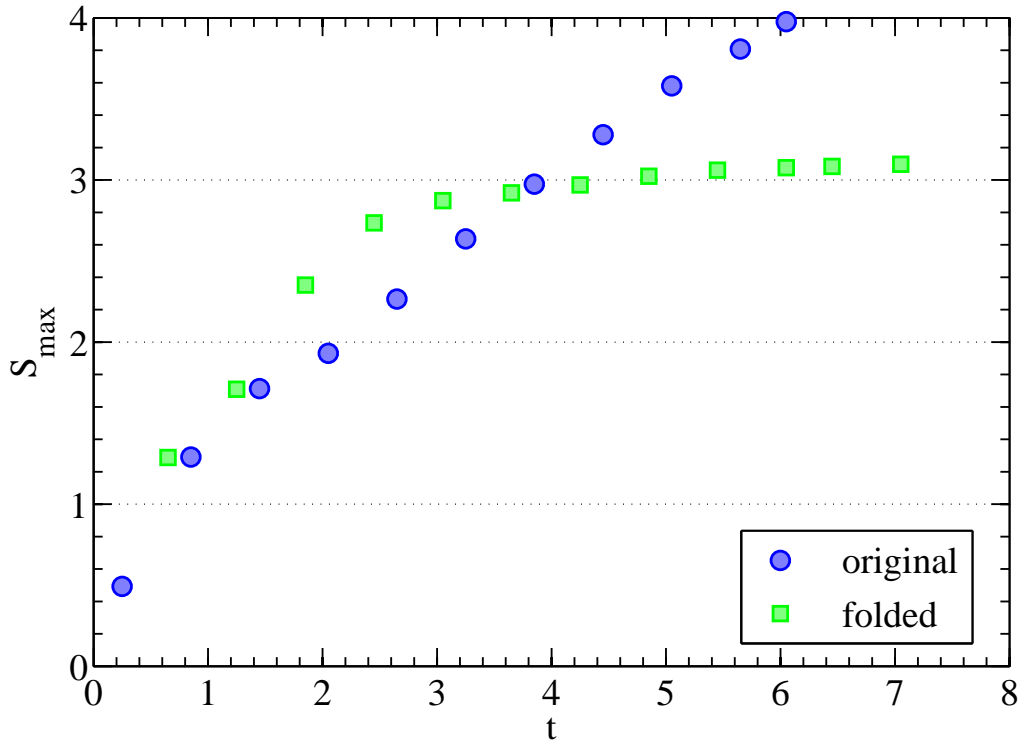}}
\caption{ Maximal entropy in the transverse dominant eigenvectors as a
  function of time in the evolution of the product state $|Z-\rangle$
  under the exactly solvable XY model \eqref{eq:XY} with parameters
  $J_x=1.5$, $J_y=0.5$ and $J_z=0$.  The simulations were run using a
  first order Trotter expansion and maximum bond dimension $D=90$
  for the dominant right eigenvectors in the original network (green triangles) and the folded one (blue circles).}
\label{fig:entropyXYprod}
\end{figure}

Another instance of a free-fermion model is the XY chain of \eqref{eq:XY}. 
In~\cite{fagotti08xy} the entanglement entropy of a block was analytically computed
and it was shown that, as for the Ising model, it can grow linearly with time after a global quench.
It is then interesting to check how accurately this scenario connects to the free-propagating picture.  
We have therefore explored the transverse entanglement in this model for what should be 
the closest scenario to the ideal case of freely propagating
excitations, i.e. the evolution of the Fourier fermionic vacuum.  The
initial state corresponds now to the product of all spins aligned
along the negative $\hat{z}$ direction, $|Z-\rangle\equiv 
\left [|1\rangle\right]^{\otimes N}$.  \Fref{fig:entropyXYprod} shows the compared maximal entropies
for transverse and transverse-folded dominant right eigenvectors in
this case. Qualitatively, the
difference in scaling reproduces the situation for the analogous Ising
case depicted in \fref{fig:entropyProdg}, with a dramatic change of slope
for the folded case which leads to a very slow increase at long
times. This change, however, occurs at a later time, and for larger
absolute values of entropy in the case of the XY chain than for the Ising model.
Therefore, even if the transverse folded eigenvectors can be well
approximated by MPS of bound dimension at long times, the largest difference
with respect to the transverse contraction without folding is likely
to be evident only at much longer times.  
The XY Hamiltonian, however, 
does not admit such an efficient decomposition of the evolution
operator as in the case of the Ising model~\cite{murg08mpo}.
Even for the lowest Trotter order, the XY evolution requires at least
twice as many MPO per step as in the Ising case, what increases the
computational cost of the simulations in an equivalent factor.

\begin{figure}[floatfix]
  \includegraphics[width=\columnwidth]{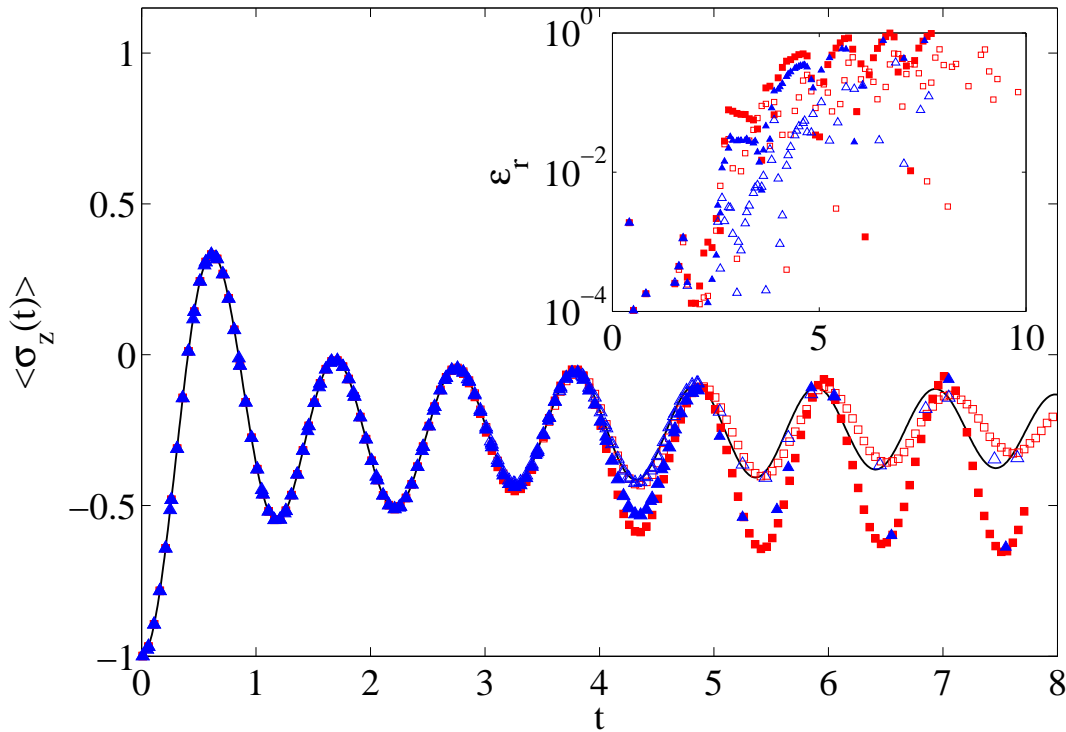}
\caption{Magnetization per site $\langle \sigma_z(t) \rangle$ as a
  function of time for the evolution, under the XY model with
  $J_x=-1$, $J_y=0.5$, $J_z=0$, of the initial product state
  $|Z-\rangle$.  The solid line is the exact result, and the solid
  (empty) symbols show the results obtained with the transverse
  (folded) contraction using bond dimension $D=60$ (red squares) and
  $90$ (blue triangles) for the eigenvectors.  The inset shows the
  relative error.}
\label{fig:polarizationXY}
\end{figure}

\Fref{fig:polarizationXY} shows the results for the
time-dependent polarization per site, $\langle \sigma_z(t)\rangle$,
when evolving the product state $|Z-\rangle$ under the XY Hamiltonian.
The folded strategy with bond dimension $D=90$ 
attains an accuracy of the order of $10^{-2}$ at times $t \simeq 6$,
which compares positively to results obtained with a light-cone
based method for a similar model
in~\cite{hastings08lightcone}.~\footnote{Times
  in~\cite{hastings08lightcone} have to be divided by a factor 4 to be
  comparable to our definition of $H_{XY}$. With this relation, the
  light-cone method was able to simulate $t\simeq 5$.}  Moreover, we
can observe the different behaviour of errors before and after folding.
The unfolded contraction deviates clearly from the exact curve after
$t\simeq 4$ and increasing the bond dimension from $D=60$ to $90$ does
not significantly improve the error. The folded version, instead,
seems to oscillate around the proper mean value, albeit somehow
slower. Increasing the bond dimension from $D=60$ to $90$ in this case
has the effect of bringing the oscillations closer to the exact
pattern, corroborating the smoother onset of errors which was already
detected in~\cite{banuls09fold}.

\subsection{\label{sec:corr}Dynamical correlators}

Often, specially relevant observables are given by another kind of dynamical
quantities, namely time-dependent two-point correlation functions,
$\langle O_{2}^{[i_2]}(t_2) O_{1}^{[i_1]}(t_1)\rangle $, computed in
the ground state of the system, and
it is then desirable that the numerical methods are able to evaluate them.

\begin{figure}[floatfix]
\begin{minipage}[t]{.35\columnwidth}
\subfigure[2D network representing the two-point correlator
  $C_{12}(i_1\ t_1,i_2\ t_2)$ in the general case.]{
 \label{fig:correl-a}
\psfrag{A}[l][l]{$t_2-t_1$}
\psfrag{B}[l][l]{$t_1$}
\psfrag{C}[bc][bc]{$i_1$}
\psfrag{D}[bc][bc]{$i_2$}
  \includegraphics[height=.9\columnwidth]{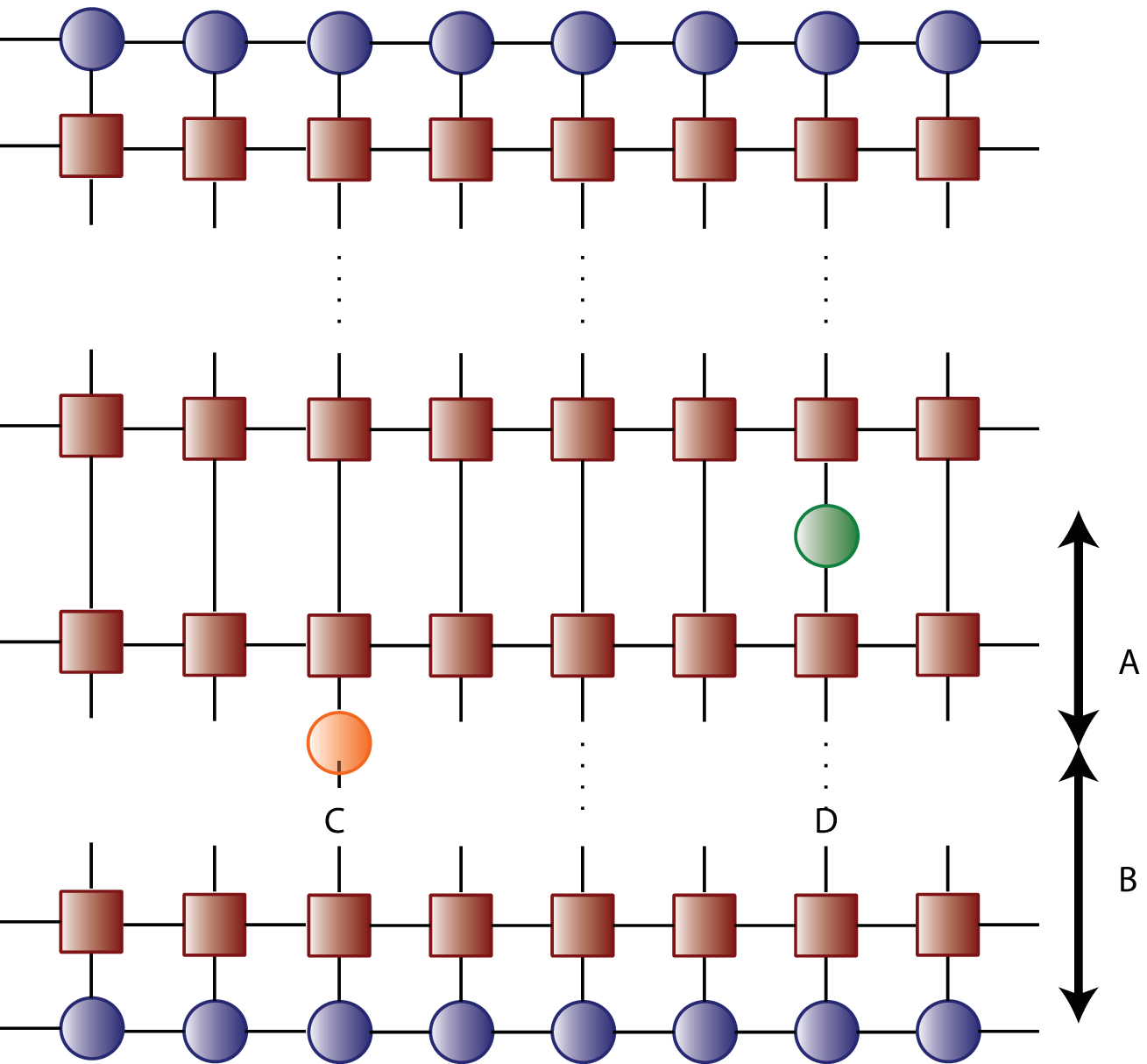}
} 
\end{minipage}
\hspace{.1\columnwidth}
\begin{minipage}[t]{.35\columnwidth}
\subfigure[Transverse  contraction reduces to a finite network.]{
 \label{fig:correl-b}
\psfrag{C}[c]{$i_1$}
\psfrag{D}[c]{$i_2$}
\psfrag{A}[c][c]{$\langle L |$}
\psfrag{B}[c][c]{$|R\rangle$}
\includegraphics[height=.9\columnwidth]{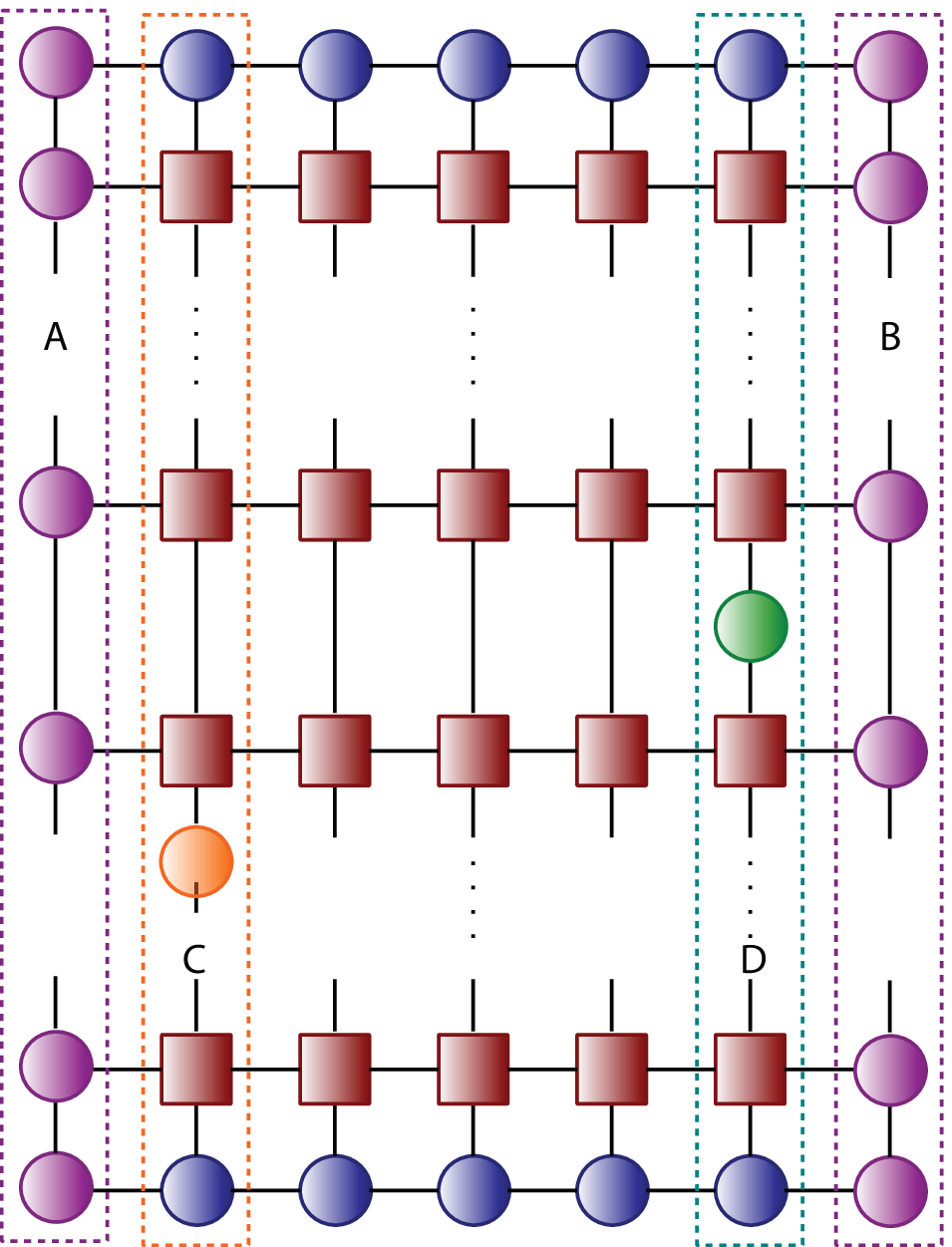}
}
\end{minipage}
\caption{
Scheme of the tensor network for the two-point time dependent
  correlator $\langle O_{2}^{[i_2]}(t_2) O_{1}^{[i_1]}(t_1)\rangle $ for $i_2>i_1$ and $t_2>t_1$.
}
\end{figure}

Even if the ground state possesses translational invariance, an
expectation value of this type breaks it, and any contraction
technique in the time direction will see its efficiency reduced,
as it will have to keep track of the causal
cone of different tensors spreading from the application of the first
operator.  The transverse techniques, on the other hand, can be easily
adapted to compute these observables with the same cost as
correlators at equal times.  

Let us consider times $t_2>t_1$, and distance $\Delta=i_2-i_1>0$.  The
two-body correlator computed in a general (not necessarily the fundamental) state $|\Psi\rangle$ can be written as 
\bea
C_{12}^{(\Psi)}(i_1\ t_1,i_2\ t_2) & \equiv & \langle
\Psi|O_2^{[i_2]}(t_2) O_1^{[i_1]}(t_1) |\Psi\rangle \nn \\ &=& \langle
\Psi | U(t_2,0)^{\dagger} O_2^{[i_2]} U(t_2,t_1) O_1^{[i_1]} U(t_1,0)
|\Psi \rangle, \nn 
\eea where $U(t_2,t_1)$ is the evolution operator
from time $t_1$ to $t_2$.  If the state $|\Psi\rangle$ is expressed as
a MPS, we can again represent this quantity by a tensor network, by approximating 
each of the evolution operators in this expression as
a product of MPO terms [\fref{fig:correl-a}].  The infinite part of the network is the same as in 
the case of a single body observable acting at time $t_2$, and can then be
reduced to the left and right dominant eigenvectors of the transfer
matrix, \beq C_{12}(i_1\ t_1,i_2\ t_2) =\frac{\langle L | E_{O_2(t_2)}
  E^{\Delta-1} E_{O_1(t_1)} | R \rangle}{\langle L | E^{\Delta+1} | R
  \rangle},
\label{eq:O(t)O(t)}
\eeq where now $E=E(t_2)$ is the transfer matrix resulting from
evolution until the longest time $t_2$, and $E_{O_i(t_i)}$ is the
column MPO operator corresponding to site $i$, on which a single-body
operator acts at time $t_i$ [see \fref{fig:correl-b}].  The cost of
finding the dominant eigenvectors is determined by the time $t_2$, as
in the case of a single-body expectation value.  The width of the
finite network left to be contracted after this step is $\Delta+3$. In
particular, if both operators act on the same site, the computation
has the same cost as that of $\langle O_2(t_2) \rangle$.\footnote{A
  similar transverse computation was done in~\cite{naef99} for the
  imaginary time Green function, in the context of transfer matrix
  DMRG.}

It is also possible to apply the folding method to the TN
in~\ref{fig:correl-a}.  From the toy model and the entropy results in
\fref{fig:entropyGSising}, we expect that for the dynamical
correlators computed in the ground state the folded eigenvectors have
a more efficient description in terms of MPS and produce better
approximations.

\begin{figure}[floatfix]
 \includegraphics[width=\columnwidth]{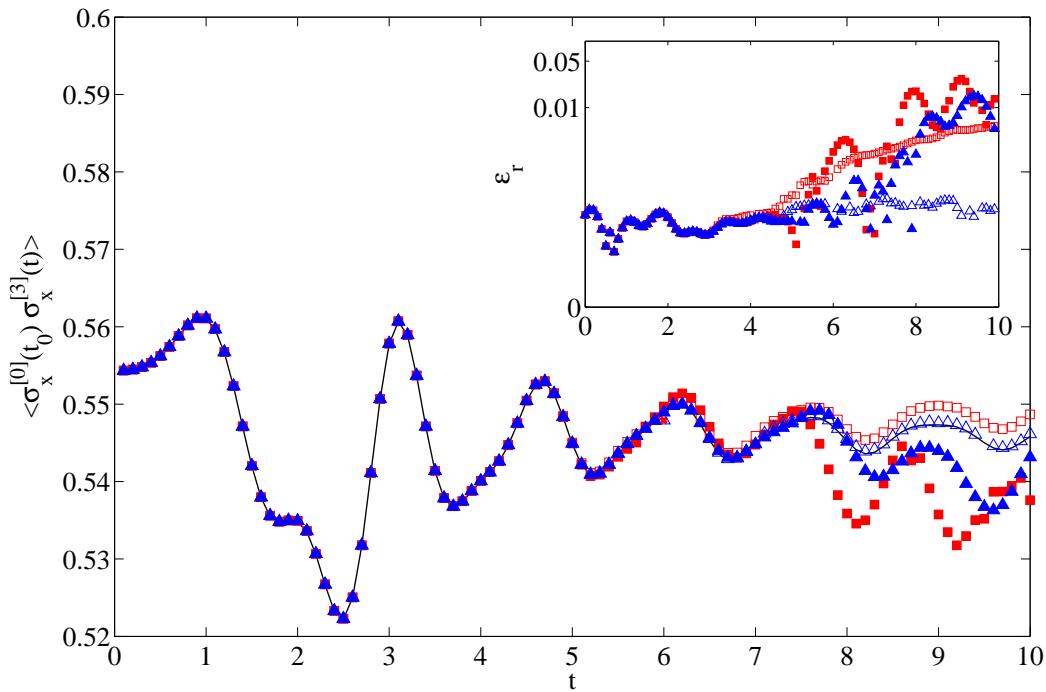}
\caption{Two-spin dynamical correlator $\langle
  \sigma_x^{[i]}(t_0)\sigma_x^{[i+\Delta]}(t_0+t) \rangle$ as a
  function of time, in the ground state of the Ising model ($g=1.1$,
  $h=0$), at distance $\Delta=3$ using $t_0=0.1$.  Shown are the exact
  result (solid line) and the results from the transverse (solid
  symbols) and folded transverse (empty symbols) contractions, using
  bond dimension $D=60$ (red squares) and $90$ (blue triangles), and a fourth order Trotter expansion with $\delta=0.1$. The
  inset shows the relative error for each series of data.  The
  approximation to the ground state, with an accuracy in energy of
  $10^{-7}$, was computed with iTEBD, using bond dimension $\xi=16$.
}
\label{fig:correlationIsing}
\end{figure}

To apply the transverse techniques to these quantities, we need to
express also the state in which the correlator is to be computed, in
this case the ground state of the model, as a tensor network. 

 We could find an appropriate TN description of the
ground state by combining real and time evolution steps in the same
TN, so that the first part of the evolution is in imaginary time, and
produces a TN approximation to the ground state and the second part,
in real time, approximates the dynamics.
Here we proceed with a simpler alternative, which consists in
first finding a MPS approximation to the ground state by means of the
iTEBD method, and then using the obtained tensor as the initial state
in our dynamical TN.

We have checked the algorithm with the computation of the two-spin
correlation function $\langle
\sigma_x^{[i]}(t_0)\sigma_x^{[i+\Delta]}(t_0+t) \rangle$ in the ground
state of the Ising model with $g=1.1$, using the transverse and
folding strategies and comparing them to the exact values. Our
results, in \fref{fig:correlationIsing}, show that the folding
strategy achieves better accuracy for the same bond dimension, $D=90$,
and is able to reach a relative error $\epsilon_r \approx 10^{-4}$
at times $t\approx 10$.  We also observe again the smoother appearance
of errors in the folded algorithm.

\begin{figure}[floatfix]
\psfrag{D}[l][l]{$t_2-t_1$}
\psfrag{e}[r][l]{}  
  \includegraphics[width=.4\columnwidth]{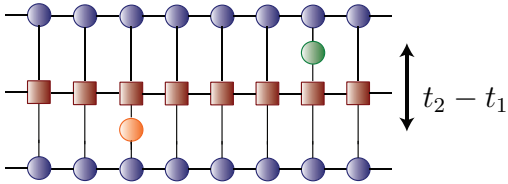}
\caption{Minimal TN for the calculation of dynamical two-point
  correlators in the ground state, to which the network in
  \fref{fig:correl-a} is reduced if the MPS initial state is the
  ground state of the evolving Hamiltonian.}
  \label{fig:minimalTN}
\end{figure}

The combined use of iTEBD and the transverse technique provides us
with another, yet more efficient way of computing the dynamical
correlators \eqref{eq:O(t)O(t)} in the ground state.  Indeed, if
$|\Psi\rangle=|\phi_0\rangle$ is the ground state of the Hamiltonian,
with energy $E_0$, the correlator 
can be simplified, $C_{12}^{(\phi)}(i_1\ t_1,i_2\ t_2) = e^{i
  (t_2 - t_1)E_0}\langle \Psi|O_2^{[i_2]} e^{-i (t_2-t_1) H}
O_1^{[i_1]} |\Psi\rangle $.  The contraction in this expression admits
a simpler TN description than the general case (see
\fref{fig:minimalTN}), which again can be efficiently contracted in the
transverse direction.

\begin{figure}[floatfix]
  \includegraphics[width=\columnwidth]{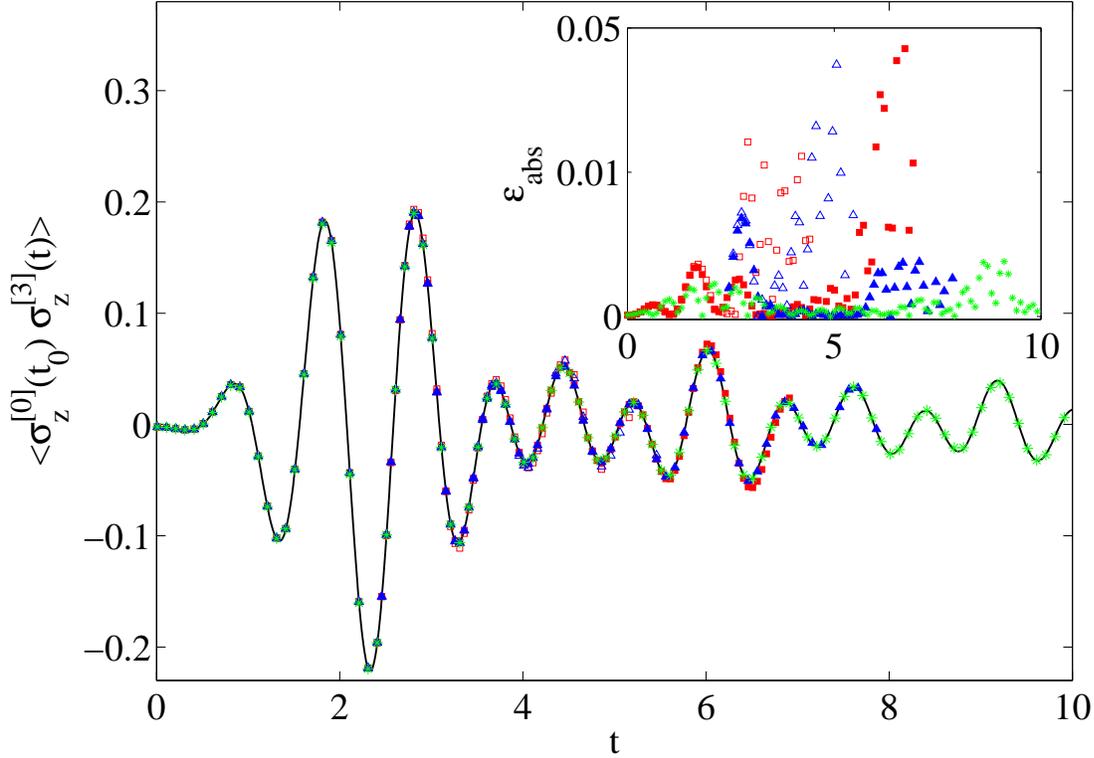}
\caption{Two-spin correlator $\langle
  \sigma_z^{[i]}(t_0)\sigma_z^{[i+\Delta]}(t_0+t) \rangle$ as a
  function of time, in the ground state of the XY model with
  $J_x=1.5$, $J_y=0.5$, $J_z=0$), at distance $\Delta=3$ using
  $t_0=0.01$.  Shown are the exact result (solid line) and the results
  from the transverse (solid symbols) and folded transverse (empty
  symbols) contractions, using bond dimension $D=60$ (red squares) and
  $90$ (blue triangles), and a first order Trotter expansion with $\delta=0.01$. The green stars show the results for the
  minimal TN transverse contraction with $D=60$. The inset shows the
  absolute error for each series.  The iTEBD approximation to the
  ground state was obtained, using bond dimension $\xi=8$, with an
  accuracy in energy of $10^{-6}$.}
\label{fig:correlationXY}
\end{figure}

\begin{figure}[floatfix]
  \includegraphics[width=\columnwidth]{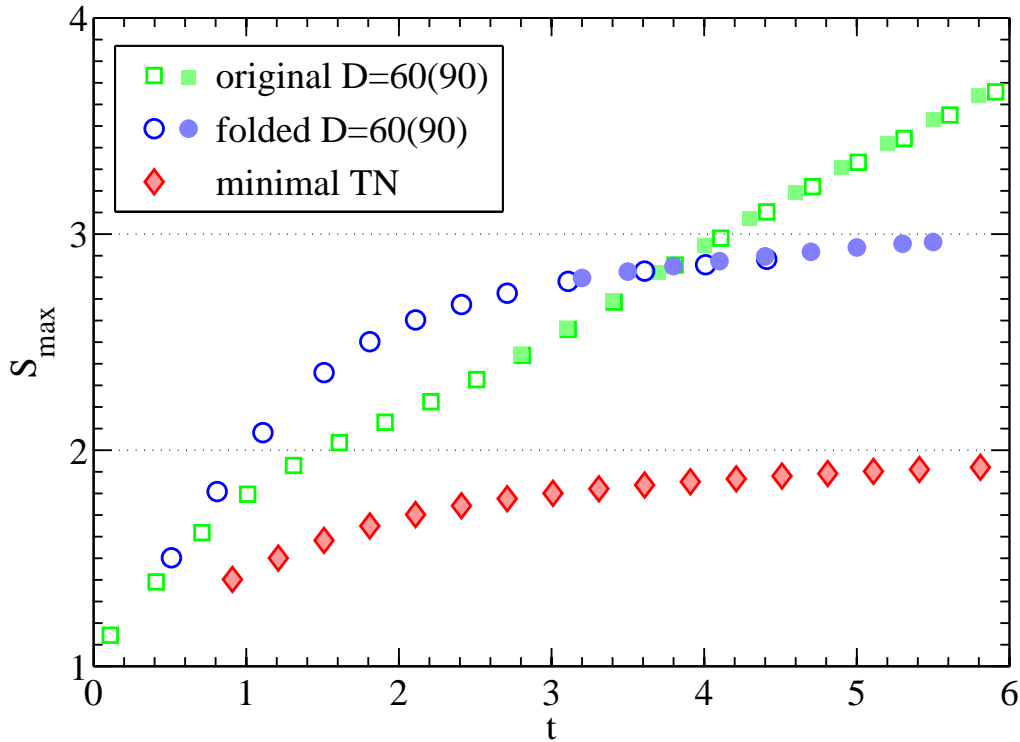}
\caption{Maximal entropy as a function of time for the transverse
  dominant eigenvectors in the tensor networks that describe the
  dynamical correlators in the XY model with $J_x=1.5$, $J_y=0.5$ and
  $J_z=0$. The ground state approximation was a MPS with bond
  dimension $\xi=8$. Shown are the results for the original transverse
  network (green squares), the folded network (blue circles) and the
  minimal TN (red diamonds).}
\label{fig:entropyXYgs}
\end{figure}

We have checked the performance of this description and compared it to
the other transverse strategies in the case of the XY model.  The
results, in \fref{fig:correlationXY}, show the clear advantage of the
minimal TN contraction, which achieves good agreement with the exact
values to times $t\approx 10$ with bond dimension as small as $D=60$.
The transverse and folded contraction of the original network were
performed with bond dimensions up to $D=90$. At the times simulated,
the contraction without folding achieves better results than after
folding. These relatively short times simulated are however not enough
to conclude whether that will still be the case at longer times. The
entropy scaling, shown in \fref{fig:entropyXYgs}, suggests the
opposite.

We can compare the maximal entropy $S_{max}$ in the dominant
eigenvectors for each of the three methods
(\fref{fig:entropyXYgs}). Qualitatively, the behaviour of the
transverse and folded strategies agrees with the previous observations
for the ground state in the Ising case (\fref{fig:entropyGSising}), with seemingly
linear increase in the case of the original TN and a much slower
growth in the case of the folded network for long times.
The maximal entropy 
for the minimal TN is much lower and also grows slower than the others,
in agreement to the much better results achieved with this technique.

\subsection{\label{sec:imaginary}Imaginary time evolution}

The basic technique described in Sect.~\ref{sec:TNconstruction} to
approximate the action of an evolution operator as a composition of
locally acting MPO can also be used for a real exponential.  Therefore
the same methods can be used to simulate imaginary time evolution of
infinite chains and to compute their ground state properties.  Since
the ground state is the limit
$$
|\Psi_0\rangle=\lim_{\tau\rightarrow\infty}
\frac{e^{- H \tau}|\Phi_0\rangle}{\|e^{- H \tau}|\Phi_0\rangle\|},
$$
we may compute the expectation value of any observable in the ground state as
\beq
\langle O \rangle_{\Psi_0}\approx\frac{\langle \Phi_0|e^{-H \tau} O e^{-H \tau}|\Phi_0\rangle}
{\langle \Phi_0|e^{-2 H \tau}|\Phi_0\rangle}
\label{eq:imag}
\eeq in the limit of large $\tau$.  This quantity can be also estimated by
 contracting a two dimensional tensor network, obtained in
this case from the application of small steps of imaginary time
evolution, $e^{-H \delta}$, approximated in terms of non-unitary MPO
factors.  The structure of the resulting network is similar, with the
transfer matrix replaced by an operator whose left and right dominant
eigenvectors, if not degenerate, are enough to reduce the network to a
finite one.  Notice that in this case, the maximum eigenvalue of the transfer matrix does not
need to be one.  It is possible to use the same techniques described
before to find a MPS approximation to the dominant eigenvectors and
compute the observables.  The transverse operators appearing in this
problem are closely related to those in the transfer matrix
DMRG approach~\cite{bursill96trans,wang97trans}.

Although the equilibrium properties of local Hamiltonians are
typically well described by the standard approach, i.e. by 
approximating the ground (or thermal) state by a MPS,
there are cases in which the transverse strategy can be advantageous.
This will be particularly true in the thermodynamic limit, in any
situation that breaks the translational invariance.

Here we describe how the transverse (folded) tensor network can be
adapted to describe two types of equilibrium observables which can
benefit from this alternative, namely the ground state properties in
the presence of impurities and the computation of thermal properties.

\subsubsection{\label{sec:impurities}Ground states of non-translationally 
invariant problems (impurities)}

The problem of finding the ground state of an infinite spin chain with
one or more impurities breaks the translational invariance and thus
poses a problem for a time-like contraction method, like iTEBD, which
will need to keep track of an increasing number of tensors (the causal
cone generated by the impurity) along the imaginary time evolution.
On the contrary, the transverse picture can deal with the problem with
similar computational cost as in the purely translationally
invariant case. 

As a test case, we considered a modification of the Ising chain in a
transverse field $H=-\left(\sum_i\sigma_z^i \sigma_z^{i+1}+g_i
\sigma_x^i \right)$, with the magnetic impurity at position $i=0$,
i.e. the magnetic field is constant over the chain, $g_i\equiv g$
$\forall i\neq 0$, except at the origin, where it takes the value
$g_0$.  This case was already analyzed in~\cite{banuls09fold}, and we
only summarize here the results for completeness.

\begin{figure}
\psfrag{i0}[c]{$i=0$}
\psfrag{ix}[c]{$i=x$}
\psfrag{langle}{$\langle L |$}
\psfrag{rangle}{$|R\rangle$}
\includegraphics[width=.4\columnwidth]{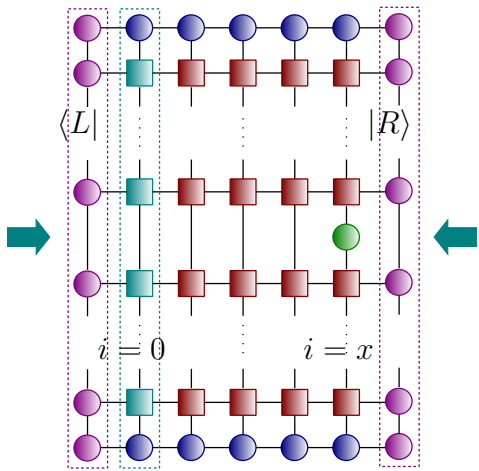}
\caption{Scheme of the tensor network for the magnetization at a
  distance $x$ of the magnetic impurity. The leftmost operator (green)
  represents the only column which is affected by the presence of the
  impurity. }
 \label{fig:impurityTN}
\end{figure}

\begin{figure}[floatfix]
  \includegraphics[width=\columnwidth]{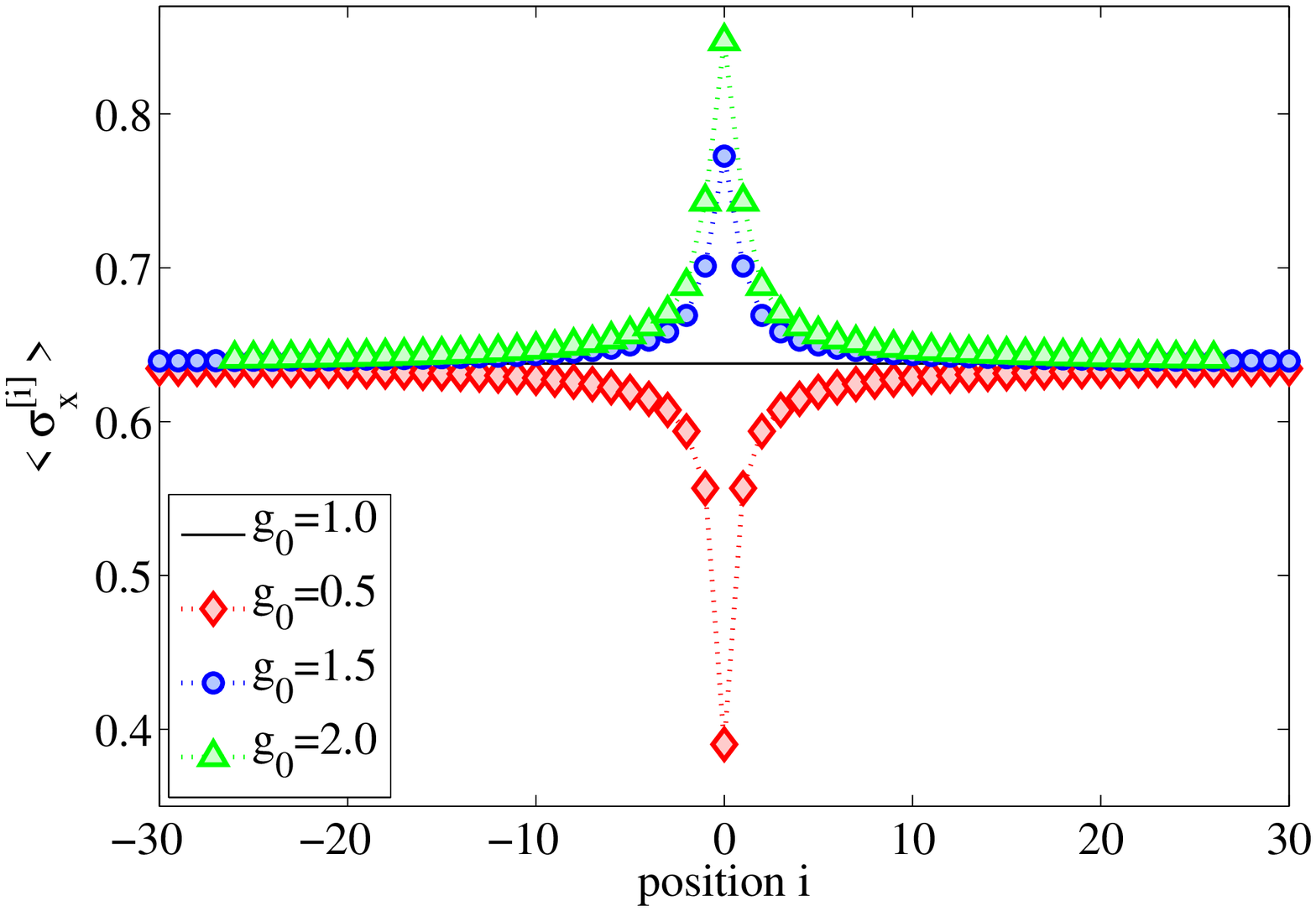}
\caption{Ground state magnetization as a function of the distance to
  the localized impurity in the Ising chain, for $g=1$ on every site
  except at the origin, where $g_0=0.5$ (red diamonds), $1.5$ (blue
  circles) and $2$ (green triangles). The solid line shows the
  magnetization in the uniform case ($g_0=1$). The results were
  obtained using bond dimension $D=40$ for the transverse
  eigenvectors, without folding the network, with imaginary time
  evolution long enough for convergence ($10^{-6}$ in the energy) of
  the critical Ising ground state.}
\label{fig:impurityMagn}
\end{figure}

The tensor network that represents the ground state obtained by
imaginary time evolution will be very similar to the translationally
invariant one obtained for uniform magnetic field. Since the impurity
affects only a localized term in the Hamiltonian, only the column
operators neighbouring $i=0$ will change.~\footnote{Depending on the
  particular decomposition of the exponentials as MPO products, the
  change may affect more than one site, but the effect will in any
  case have a bounded range, independent of the time.}  In particular,
the computation of the left and right dominant eigenvectors will not
be affected by the impurity, which will only appear in the final
contraction of the finite tensor network. The particular network to be
contracted will depend then on the observable we want to calculate.
\Fref{fig:impurityTN} pictures the resulting tensor network for
the computation of the site dependent magnetization, $\langle
\sigma_x^{[i]} \rangle$, i.e. the magnetization at distance $i$ from
the impurity.  The numerical results, in \fref{fig:impurityMagn},
illustrate the success of the algorithm at describing the screening of the impurity for different values of the
magnetic field at the origin.

\subsubsection{\label{sec:thermal}Thermal states}


The transverse folded network can be modified to represent the thermal
equilibrium state of a local Hamiltonian.  Although it is possible to
efficiently approximate such thermal state by a
MPO~\cite{hastings06solving}, and there are efficient algorithms to
find this kind of description~\cite{verstraete04mpdo}, it is worth
mentioning that the standard approach already proceeds in a folded
space, as it must work with MPO and not MPS. Therefore folding the
network does not increase the effective dimension of the sites with
respect to the standard direction of contraction.  Additionally, since the
folding contraction does not construct the MPO description of the
state, it can avoid some difficulties the standard method needs to
face, such as imposing positivity of the density matrix.  Moreover, the
transverse approach can be specially useful when the translational
invariance is broken, as could be in the localized impurity problem.
Finally, imaginary and real time evolutions can also be combined to
compute time dependent observables from the thermal state, such as
dynamical correlators at finite temperature.

\begin{figure}[floatfix]
\hspace{-.05\columnwidth}
\begin{minipage}[b]{.4\columnwidth}
\subfigure[2D network representing the unnormalized thermal state $\rho(\beta)$.]{
 \label{fig:rho-b}
\psfrag{contractL}[c][c]{contraction}
\psfrag{contractR}[c][c]{}
\psfrag{identity}[bc][bc]{$\ident$}
\psfrag{step}[bc][bc]{$\mathrm{e}^{-\delta\beta H}$}
  \includegraphics[height=.9\columnwidth]{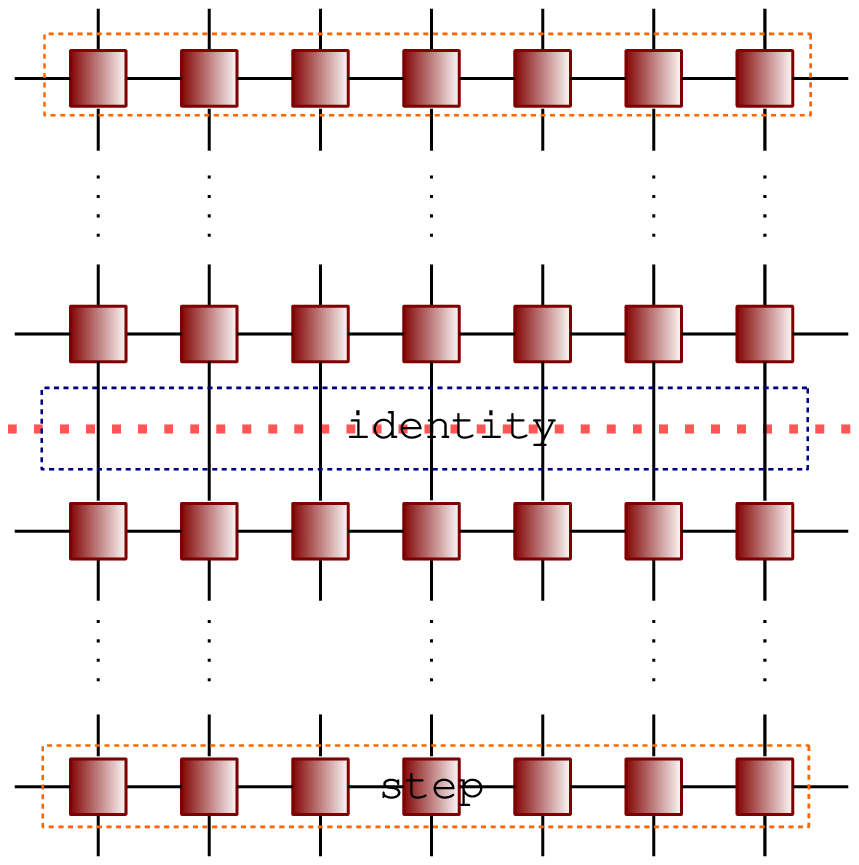}
} 
\end{minipage}
\hspace{.1\columnwidth}
\begin{minipage}[b]{.4\columnwidth}
\subfigure[Transverse  contraction of the network for a thermal expectation value.]{
 \label{fig:rho-fold}
\psfrag{op}[c][c]{$O$}
\psfrag{tra}[c][c]{\parbox{.3\columnwidth}{traced indices}}
\psfrag{langle}[c][c]{$\langle \tilde{L} |$}
\psfrag{rangle}[c][c]{$|\tilde{R}\rangle$}
\psfrag{fol}[cl][cl]{\parbox{.2\columnwidth}{folding axis}}
  \includegraphics[height=.9\columnwidth]{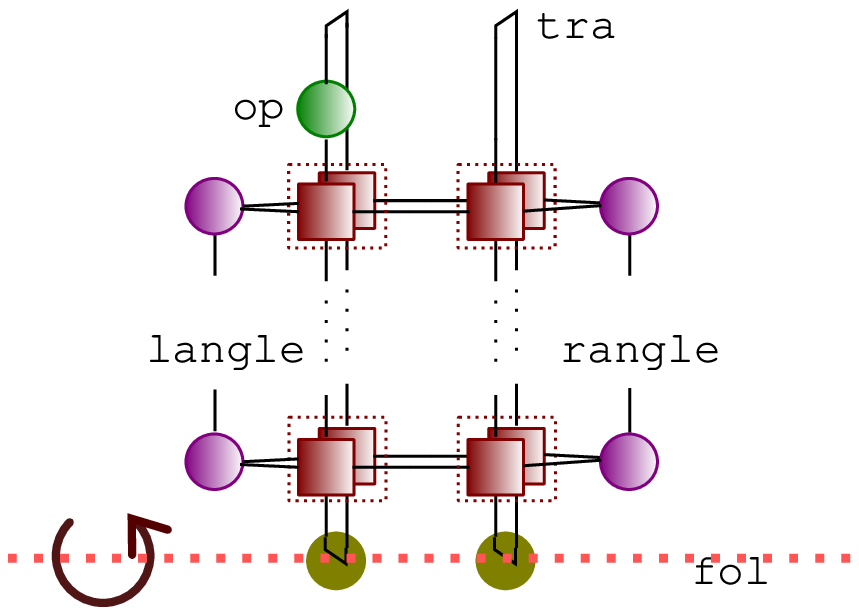}
}
\end{minipage}
\caption{Expectation values in thermal states correspond naturally to a folded network.
}
\end{figure}

The thermal density matrix at inverse temperature $\beta$, for the
Hamiltonian $H$, up to normalization, is given by the exponential 
$\rho_{th}(\beta)\propto \mathrm{e}^{-\beta H}=\mathrm{e}^{-\frac{\beta}{2}
  H} \ident \mathrm{e}^{-\frac{\beta}{2} H}$.  This is formally
identical to the evolution of the identity operator in imaginary time
${\beta}/{2}$ under the Hamiltonian $H$.  The exponential
operators in this expression can be also approximated by the same
Trotter expansion described in section~\ref{sec:TNconstruction}, so
that the density matrix can be expressed as a two dimensional tensor
network, as shown in \fref{fig:rho-b}.  To compute the expectation
value of a given local observable we need to evaluate the trace
$\mathrm{tr}[\hat{O} \rho_{th}(\beta)]$.  To this end, we apply the operator
$O$ and sum over all indices which, in the pictorial
representation of the tensor network, corresponds to connecting pairs
of open legs.  The result [see \fref{fig:rho-fold}] is a
naturally folded network, where the folding axis is taken to lie along
the initial identity operator.  This network is very similar to the
one obtained after folding for the imaginary time evolution of a pure
state, differing only in the tensors that occupy the place of the
initial state.  In fact, the thermal network is equivalent to the
asymmetric contraction $\langle\Phi|(\hat{O}
\mathrm{e}^{-\frac{\beta}{2}H}\otimes
\mathrm{e}^{-\frac{\beta}{2}\bar{H}})|\Phi\rangle$.  The network can
be thus contracted using the transverse method in a completely
analogous manner to the imaginary time evolution of a pure state after
folding, and the remaining norm factor can be computed in the same way
for the identity operator $\hat{O}=\ident$.

\begin{figure}[floatfix]
  \includegraphics[width=\columnwidth]{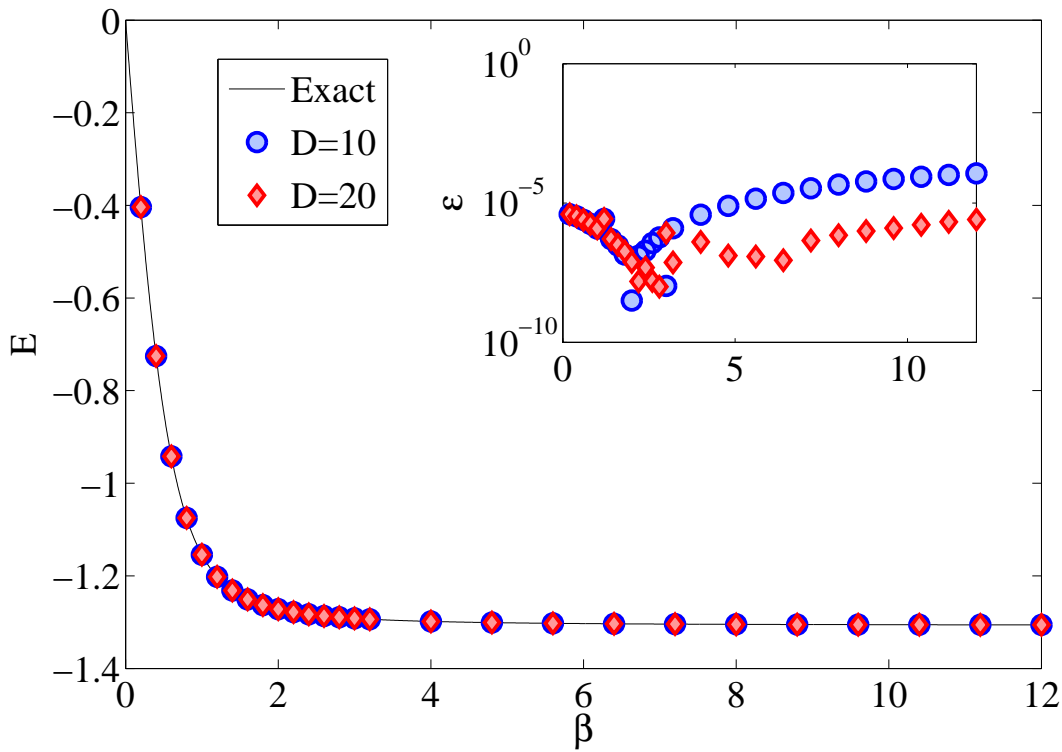}
\caption{Energy per particle as a function of the inverse temperature
  $\beta$ in the thermal state for the integrable Ising model with
  $g=1.05$.  The plot shows the exact result (solid line) and the
  energy computed with the folding method for bond dimension 10
  (blue circles) and 20 (red stars), using a fourth order Trotter expansion
  with step $\delta=0.1$.  The inset shows the relative error.  }
\label{fig:thermIsing}
\end{figure}

\begin{figure}[floatfix]
  \includegraphics[width=\columnwidth]{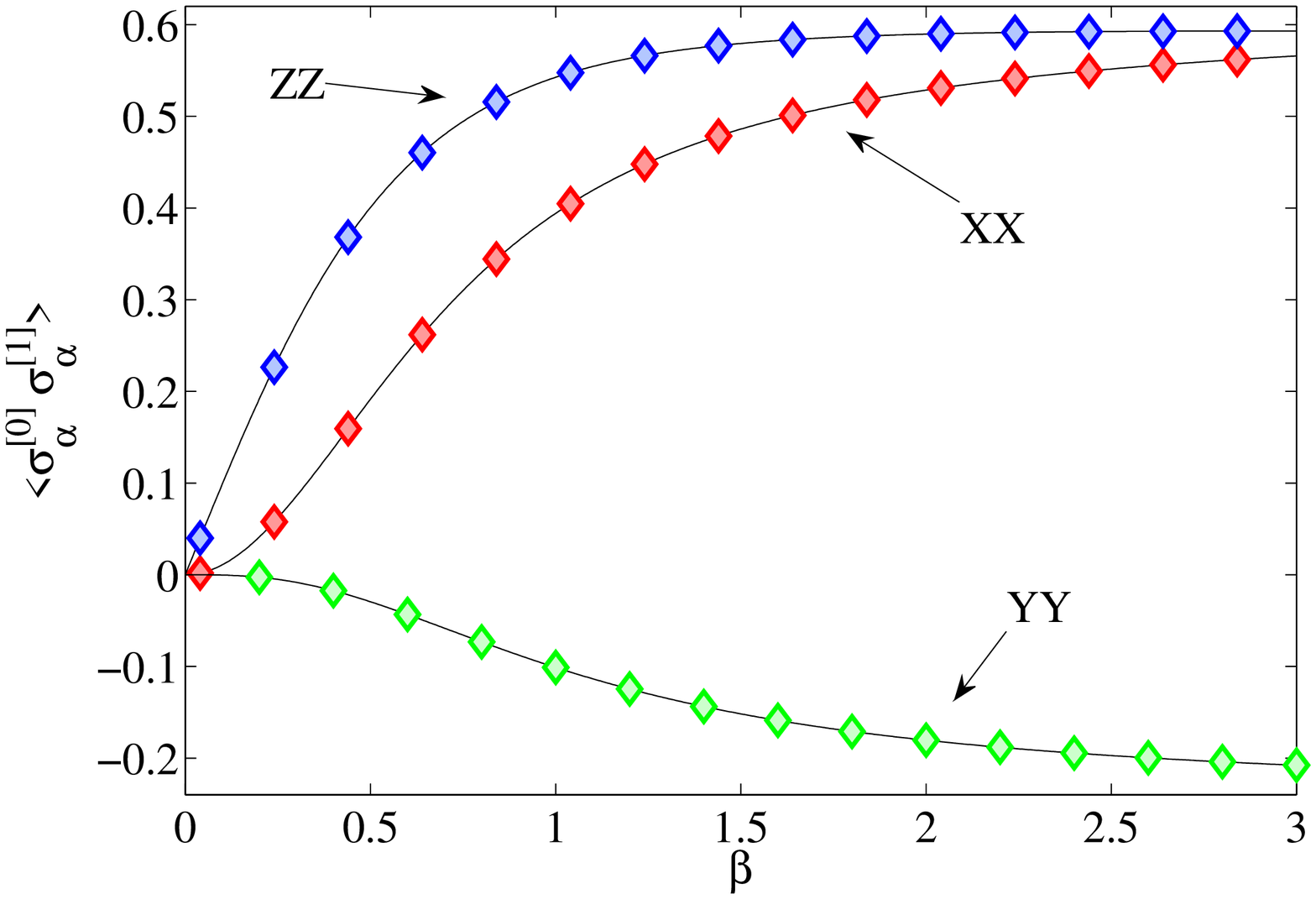}
\caption{Nearest-neighbour correlations in the thermal state as a
  function of the inverse temperature $\beta$ in the thermal state for
  the integrable Ising model with $g=1.05$.  The plot shows the exact
  result (solid line) for three types of spin-spin correlations, and
  the numerical results (diamonds) obtained with the folding method for bond
  dimension 10, using a fourth order Trotter expansion with step
  $\delta=0.1$.}
\label{fig:thermCorrelIsing}
\end{figure}

We checked the accuracy of thermal expectation values computed with
this method using the exactly solvable Ising model as a benchmark. As
shown in the figures, very accurate results are obtained with a modest
bond dimension.  A relative error in the energy below $\varepsilon=10^{-5}$ is
achieved for bond dimension $D=20$ (\fref{fig:thermIsing}). In
\fref{fig:thermCorrelIsing} we show also the good convergence of the
spin-spin correlations with a bond dimension of only $D=10$ for the
transverse eigenvectors.

\section{\label{sec:conclu}Discussion}

In this work we have analyzed in some depth the properties of the
transverse folding method, recently introduced for the dynamical
simulation of infinite quantum spin chains.  The folding method
restates the problem of computing time dependent observables as 
 a two dimensional tensor network contraction, in which the second
dimension corresponds to time.  Some of the previously existing
algorithms for simulating time evolution can also be reinterpreted as
different strategies for the contraction of this network.  The
performance of the various methods is then related to the entanglement
structures that appear in the network, depending on its boundaries,
which include the initial state, and on the direction chosen to
perform the contraction.

We have presented a dynamical model defined by a tensor network that,
despite its simplicity, exhibits the main characteristics of the
contraction that depend on the direction and the initial state.  In
particular, the scaling of entanglement with time can be very
different depending on the contraction strategy, thus deciding the
success or failure of one approach or another.  Notably, some of the
extreme cases in which one technique overpowers the other can be
easily understood from the features of the toy model.

We have also studied the reduction of transverse entanglement achieved
by folding in the case of a real Hamiltonian evolution, under a family
of models of the Ising type. We have found that the folding strategy
often results in a more favourable scaling of the entanglement with
time, even in some cases that are far from the ideal picture of
free-propagating excitations that motivates this approach.

We can conclude that, although in most problems the time extent that
can be simulated by these methods will remain limited, understanding
the entanglement contents in the state and also in the time direction
can be determinant to decide on the feasibility of a given simulation,
and to choose the most convenient simulation method for a
particular scenario.

As we have discussed, transverse contraction strategies, including the folding of the TN,
can be applied to a variety of physical problems involving infinite
chains. In particular, to situations that break translational
invariance, as can be the presence of impurities, and to the
computation of two-body correlations at different times.

\acknowledgments
We thank M. Hastings and F. Verstraete for discussions. 
This work was partly funded by EU through QUEVADIS, and by
DFG through the SFB
631, and the DFG Forschergruppe 635.

\bibliography{foldingL}
\end{document}